\documentclass[onecolumn,12pt,draftcls]{IEEEtran}
\IEEEoverridecommandlockouts

\usepackage{amsmath}
\usepackage{amsfonts}
\usepackage{bbding}
\usepackage{amssymb}
\usepackage{array}

\usepackage{graphicx}
\usepackage[named]{algo}
\usepackage{algorithmic}
\usepackage[compress]{cite}
\makeatletter
\renewcommand{\citepunct}{,\penalty\@m\hskip.13emplus.1emminus.1em}
\renewcommand{\citedash}{\hbox{--}\penalty\@m}
\makeatother
\usepackage{color}
\allowdisplaybreaks

\usepackage{amsthm}
\usepackage{stfloats}

\newtheorem{pro}{Property}
\newtheorem{prop}{Proposition}

\begin{document}
\title{Deep Learning for Hybrid 5G Services in Mobile Edge Computing Systems: Learn from a Digital Twin}

\author{
	\IEEEauthorblockN{Rui Dong, Changyang She, Wibowo Hardjawana, Yonghui Li, \\ and Branka Vucetic}

	\thanks{The authors are with the School of Electrical and Information Engineering, University of Sydney, Sydney, NSW 2006, Australia (email: \{rui.dong, changyang.she, wibowo.hardjawana, yonghui.li, branka.vucetic\}@sydney.edu.au).}

}

\maketitle

\begin{abstract}
In this work, we consider a mobile edge computing system with both ultra-reliable and low-latency communications services and delay tolerant services. We aim to minimize the normalized energy consumption, defined as the {energy consumption per bit}, by optimizing user association, resource allocation, and offloading probabilities subject to the quality-of-service requirements. The user association is managed by the mobility management entity (MME), while resource allocation and offloading probabilities are determined by each access point (AP). We propose a deep learning {(DL)} architecture, where a digital twin of the real network environment is used to train the {DL} algorithm off-line at a central server. From the pre-trained deep neural network (DNN), the MME can obtain user association scheme in a real-time manner. Considering that real networks are not static, the {digital twin monitors the variation of real networks and updates the DNN accordingly}. For a given user association scheme, we propose {an optimization} algorithm to find the optimal resource allocation and offloading probabilities at each AP. Simulation results show that our method can achieve {lower normalized energy consumption with less computation complexity} compared with {an existing method} and approach to the performance of the global optimal solution.
\end{abstract}

%


\section{Introduction}
\subsection{Backgrounds and Motivations}
In the 5th generation (5G) communication systems, there are diverse applications ranging from high data rate delay tolerant services to ultra-reliable and low-latency communications (URLLC) \cite{3GPP2017Scenarios}. {By achieving ultra-low end-to-end (E2E) delay and ultra-high reliability, URLLC lies the foundation for emerging latency-critical applications,} such as factory automation, autonomous vehicles, and virtual/augmented reality \cite{Philipp2017IoT}. {Devices in these applications will generate some tasks that require to be processed within a short time. To reduce processing time at the local server of each device and to avoid delays in backhauls and core networks, mobile edge computing (MEC) is one promising solution \cite{Yuyi2017MEC}.} However, when a task is packetized in a short packet and offloaded to a MEC server via a wireless link, the packet may be lost when the channel is in deep fading \cite{Changyang2018TWC}. Besides, short blocklength channel codes will cause the none-zero decoding error rate, even for an arbitrarily high signal-to-noise ratio (SNR) \cite{Yury2014Quasi}. Thus, achieving ultra-high reliability and ultra-low latency is very challenging in MEC systems.

On the other hand, mobile devices have only limited battery capacities. Improving the battery lifetime or energy efficiency (EE) of users is an urgent task \cite{Yuyi2017MEC,Kang2018Energy,Zhang2018Energy}. By offloading tasks to MEC servers, we can save energy consumptions at the local servers (equipped at mobile devices), but extra energy is consumed for data transmissions. To minimize the total energy consumption of each user, we need to optimize the offloading probability. In a MEC network with multiple MEC servers and multiple users, the problem that optimizes user association, resource allocation, and offloading probabilities is non-convex and complicated. How to improve EE by solving a non-convex problem in the scenario with both URLLC services and delay tolerant services remains an open problem.

To find the optimal solution to the problem, there are two kinds of approaches: optimization algorithms and machine learning algorithms. Since optimization algorithms need to search the optimal solution when channels change, they are suitable for small-scale problems, such as resource allocation in a single access point (AP) scenario \cite{Zhang2018Energy,you2018asynchronous}. When the scale of the problem grows, deep learning (DL) algorithms have the potential to find a near optimal solution in a real-time manner \cite{Guo2018Access}. {For example, a deep neural network (DNN) can be used as an approximator of an optimal resource allocation policy \cite{Guo2018Access}. The resource allocation obtained from the optimal policy can be used as labeled samples to train the DNN.}
Once the training of the DNN is finished, we can compute the resource allocation from it with different channel realizations.

{To train a DL algorithm, we first need to obtain optimal policies from simplified system models. However, optimal policies may not be available in practical systems. Thus, some other techniques are needed to enable DL algorithms. One approach that does not require labeled training samples is deep reinforcement learning \cite{luong2018applications}. By learning from the feedback of real-environment,  deep reinforcement learning is widely used to maximize long-term rewards of Markov decision processes. This approach is not suitable for URLLC services due to the following two reasons. First, maximizing the long-term reward cannot guarantee the delay and reliability requirements in each time slot. Second,} to check whether the packet loss probability satisfies the reliability requirement from the feedback of real network environment, a user needs to transmit a large number of packets. If the required packet loss probability of $10^{-7}$, a user needs to transmit more than $10^7$ packets, which may be larger than the total number of packets that will be generated within the service time of the user. To handle this issue, we need to compute the packet loss probability with the help of theoretical results that are obtained with model-based methods.

To merge the model-free deep learning algorithms with model-based theoretical results, we establish a digital twin of the real network environment. As shown in \cite{GE2017}, a digital twin is a virtual digital model of the real network that consists of data from the real network {(e.g., network topologies, schedulers, and channels)} and fundamental rules from theoretical studies {(e.g., tradeoffs in information and queueing theories)}. With the help of a digital twin, we can compute the energy consumption, delay, and packet loss probability of a certain decision on user association and resource allocation. {In addition, by monitoring the variations of the real network environment, the system can update the digital twin for training the DNN. As such, it is possible to implement deep learning algorithms in non-stationary environment.} Nevertheless, how to apply a digital twin in the {DL} architecture for hybrid 5G services in MEC systems remains unclear.

Motivated by the above issues, we will answer the following questions in this paper: {1) how to improve EE for URLLC and delay tolerant services in MEC systems? 2) How to establish the digital twin that mirrors the real network environment? 3) How to establish a DL framework based on the digital twin to solve non-convex optimization problems?}


\vspace{-0.2cm}
\subsection{Our Solutions and Contributions}
In this paper, we would like to improve EE of users in a MEC system, subject to the delay and reliability constraints of URLLC services and the stability constraint of delay tolerant services. A digital twin of the real network is adopted to train the DL algorithm. To the authors' knowledge, this is the first paper that incorporating the concept of the digital twin with wireless networks. The main contributions of this paper are summarized as follows.
\begin{itemize}
\item We propose {a digital twin enabled DL} framework for improving EE of URLLC and delay tolerant services in a MEC system with multiple APs. The normalized energy consumption, defined as {the energy consumption per bit}, is minimized by optimizing user association, resource allocation, and offloading probabilities. {With this framework, the optimal user association scheme is first explored in the DL framework, and then approximated by a DNN, which is first trained off-line at the central server and then sent the mobility management entity (MME). After the training phase, the central server keeps updating the DNN according to the variation of the real network environment.} For a given user association scheme, each AP optimizes the resource allocation and task offloading policy.
\item {We establish the digital twin of the MEC system, where the network topology, the channel and queueing models, and the fundamental rules are adopted to mirror the real system. In addition, the behavior of each AP, i.e., the optimal resource allocation and task offloading policy, is included in the digital twin. Although the problem at each AP is non-convex, we propose an algorithm that converges to the global optimal solution with linear complexity.}
\item {We design exploration policies that generate multiple user association schemes in each learning epoch. For each user association scheme, the normalized energy consumption, delay, and reliability are evaluated in the digital twin. Then, the best one is saved in the memory as labeled training samples. To compare the efficiency of different exploration policies, the normalized energy efficiency and computing complexity of different exploration policies are illustrated with simulation results.}
\end{itemize}
{Furthermore, simulation results show that the proposed DL framework can achieve a lower normalized energy consumption with less computing complexity compared with an existing solution, and can approach to the global optimal solution.}
\vspace{-0.2cm}

\section{Related Work}
How to improve EE of mobile devices in MEC systems subject to the delay constraint has been widely studied in existing literature \cite{you2018asynchronous,Zhang2018Energy,cheng2018energy,ko2018wireless,Jizhe2019Joint}. {To study the tradeoff between EE and latency,} a weighted sum of energy consumption and latency was minimized in a single-AP scenario \cite{Zhang2018Energy}. EE was maximized subject to the delay constraint in single-AP scenarios and multi-AP scenarios in \cite{you2018asynchronous} and  \cite{cheng2018energy}, respectively. The authors of \cite{ko2018wireless} analyzed the EE and latency with stochastic geometry and provided some useful guidelines to network provision and planning. {The above studies mainly focused on one kind of services, and neglected the heterogeneities of services. To address this issue, a game theory approach was proposed in \cite{Jizhe2019Joint}, where resource management and user association were optimized in multi-access MEC systems.}

How to apply machine learning algorithms for user association or task offloading in MEC systems was also studied in some recent works  \cite{chen2018performance,min2017learning,xu2017online,Angela2018deep}. Deep Q-learning was used to minimize the task execution cost by optimizing offloading decision according to channel state information, queue state information, and energy queue state of the energy harvesting system \cite{chen2018performance}. A similar method was also applied for energy harvesting of IoT devices in \cite{min2017learning}. The authors of \cite{xu2017online} proposed an efficient reinforcement
learning-based resource management algorithm to incorporate renewable energy into MEC systems. More recently, a deep reinforcement learning framework for task offloading was studied in a single-AP scenario \cite{Angela2018deep}.

The above studies provided useful insights and promising machine learning algorithms in MEC systems, but they did not consider 5G services. Supporting URLLC in MEC systems was studied in \cite{Liu2017MEC} and \cite{liu2018offloading}. In \cite{Liu2017MEC}, the long-term average power consumption of mobile devices is minimized subject to the latency and reliability constraints. The weighted sum of delay and reliability is minimized in \cite{liu2018offloading} for a single-user. Nevertheless, how to serve hybrid 5G services in MEC systems remains unclear, and deserves further study.

\vspace{-0.2cm}
\begin{table}[htb]
	\centering
	\caption{{Notations}}
	\label{notations}
\vspace{-0.2cm}	
	\scalebox{0.8}{
		\begin{tabular}{|l|l||l|l|}
			\hline
			Notation & Definition  & Notation & Definition \\ \hline
			$x$ &  scalar               & 	$\textbf{x}$&   vector  \\ \hline
			$\mathbb{E}$  & expectation &   $(\cdot)^{\rm T}$ & transpose operator \\ \hline
			$M$ &  number of APs & $\xi=\{\rm u, b\}$ & superscript representing URLLC and delay tolerant services \\ \hline
			$K^{\xi}$, $\mathcal{K}^{\xi}$&   number of users and set of users & $T_s$  &  duration of each slot  \\ \hline
			$S_m$  &  service rate of the $m$th MEC server    & $C_k^\xi$  &  service rate of the $k$th user    \\ \hline
			$C_k^{\max,\xi}$ &  maximum computing capacity of the $k$th user  & $\lambda_{k}^\xi$ & average task arrival rate generated by the $k$th user\\ \hline
			$b_k^\xi$ & number of bits of each task & $c_k^\xi$ & number of CPU cycles required to process each task \\ \hline
			$\beta_{m,k}^\xi = \{0,1\}$  &  user association indicator &	$N_{m,k}^\xi$ & number of allocated subcarriers \\ \hline
			$W$ & bandwidth of each subcarrier &	$\alpha_{m,k}^{\xi}$ & large-scale channel gain\\ \hline
			$g_{m,k}^{\xi}$ & small-scale channel gain&	$P_k^{\rm t,{\xi}}$ & transmit power\\ \hline
			$\Phi$ & SNR loss coefficient&	$N_0$ & single-side noise spectral density\\ \hline
			$f_Q^{-1}$ & inverse of Q-function&	$\varepsilon_k^{\rm d,{\rm u}}$ & decoding error probability\\ \hline
			$x_k^\xi$   &  offloading probability  &	$e_k^{\rm loc, \xi}$   &  energy consumption per CPU cycle  \\ \hline
			$E_k^{\rm loc, \xi}$   &  energy consumption per packet at the local server  &	$D_k^{\rm lc, u}$   &  processing delay on local server  \\ \hline
			$D_k^{\rm lq, u}$   &  queueing delay on local server  &	$D^{\rm max, u}$   & maximum delay \\ \hline
			$\epsilon_k^{\rm lq,u}$   & queueing delay violation probability in local server  &	$\epsilon^{\rm max,u}$   & maximum queueing delay violation probability \\ \hline
			$\epsilon_k^{\rm mc,u}$   & processing delay violation probability in MEC server  &	$\rho_m^{\rm mc}$   & workload of the MEC server  \\ \hline
			$\bar{c}_k^{\rm b}$ & average number of required CPU cycles  &	$\bar{b}_k^{\rm b}$ & average number of bits in a packet  \\ \hline
			$D_k^{\rm mc, u}$   & processing delay on MEC server   &	$ \eta _k^\xi$ & normalized energy consumption  \\ \hline
			$N^{\max}$  & total number of subcarriers   &	$ P_k^{\max, \xi} $   &  maximal transmit power   \\ \hline
			$\hat{\boldsymbol{{\beta}}}$ & direct output of the DNN &	$\tilde{\boldsymbol{\beta}}$ & the best user association scheme obtained from the digital twin\\ \hline
		\end{tabular}
	}
\end{table}

\section{System Model}
\subsection{MEC System}
\begin{figure}[htbp]
	\vspace{-0.2cm}
	\centering
	\begin{minipage}[t]{0.7\textwidth}
		\includegraphics[width=1\textwidth]{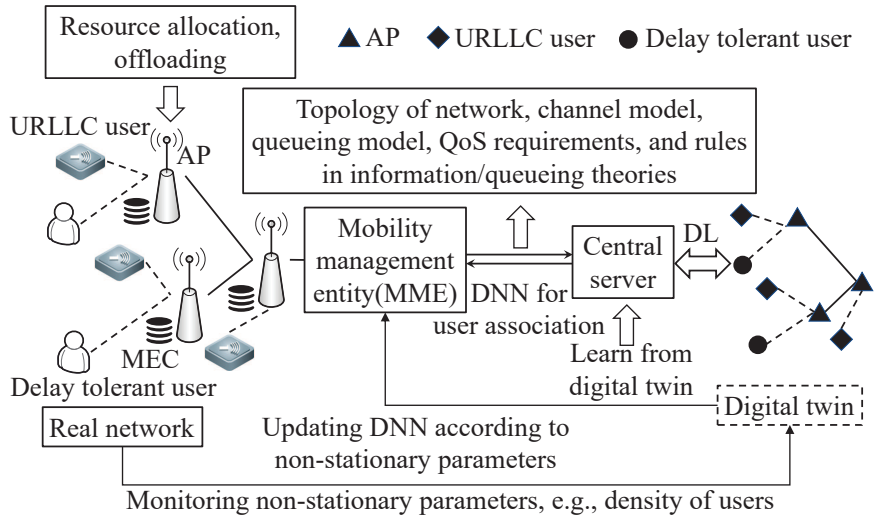}
	\end{minipage}
	\vspace{-0.5cm}
	\caption{{System model.}}
	\label{sys}
	\vspace{-0.2cm}
\end{figure}

We consider a MEC system shown in Fig. \ref{sys}, where $M$ APs serve $K^{\rm u}$ URLLC services and $K^{\rm b}$ delay tolerant services, which are indexed by $\mathcal{K}^{\rm u} = \{1,...,K^{\rm u}\}$ and $\mathcal{K}^{\rm b}= \{K^{\rm u}+1,...,K^{\rm u}+K^{\rm b}\}$, respectively. For notational simplicity, we use a superscript $\xi = \{\rm u, \rm b\}$ to represent the types of services in this work. If $\xi = {\rm u}$, a parameter is used in URLLC services. Otherwise, it is used in delay tolerant services. {All the notations used in this paper are listed in Table \ref{notations}.}

The APs are connected to the MME that is in charge of user association. To establish the digital twin, the MME sends some parameters and models of the network to a central server, where the user association scheme is optimized by a DL algorithm that learns from the digital twin. After the training phase, a DNN for user association is sent to the MME. With a given user association scheme, the network can be decomposed into single-AP problems.\footnote{Frequency reuse factor is less than one that different bandwidth is allocated to adjacent APs. As such, there is no strong interference, and weak interferences are considered as noise.} For each single-AP problem, the AP optimizes resource allocation and task offloading for users that are associated with it.

Each AP is equipped with a MEC server and each user has a local server. Time is discretized into slots. The duration of each slot is $T_{\rm s}$. The service rates of the $m$th MEC server and the $k$th user are denoted as $S_m$ (CPU cycles/slot) and $C_k^\xi$ (CPU cycles/slot), respectively. The $k$th user can adjust $C_k^\xi$ within the regime $[0,C_k^{\max,\xi}]$, where $C_k^{\max,\xi}$ is the maximum computing capacity of the user.

{Without loss of the generality, non-stationary parameters in a system can be classified into two categories. The first category of parameters is highly dynamic, such as the large-scale channel gains and the average task arrival rates. The other category of parameters varies slowly, such as the density of users in a certain area. For the first category of parameters, we include them in the input of the DNN. For the second category of parameters, the system monitors their values and updates them in the digital twin. Then, the DNN learns from the updated digital twin. Rather than training a new DNN, the previous well-trained DNN will be used to initialize the new DNN. In this way, the output of the DNN changes with non-stationary parameters.}

\subsection{Computation Tasks and Communication Packets}
The computation tasks of the $k$th user are characterized by $(\lambda_{k}^\xi,b_k^\xi, c_k^\xi)$, where $\lambda_{k}^\xi$ (packets/slot) is the average task {arrival} rate generated by the $k$th user, $b_k^\xi$ (bits/packet) is the number of bits of each task (i.e., the size of a packet), and $c_k^\xi$ (cycles/packet) is the number of CPU cycles required to process each task. We assume that each task is conveyed in one packet, and the relation between $b_k^\xi$ and $c_k^\xi$ is given by $c_k^\xi = k_1 b_k^\xi$, where $k_1 > 0$ (cycles/bit) depends on the computational complexity of the task \cite{miettinen2010energy,munoz2015optimization,wang2016mobile}.

For URLLC services, we assume that the packet size and the number of CPU cycles required to process each packet are constant (e.g., $32$~bytes \cite{3GPP2017Scenarios}), and the packet arrival process follows a Bernoulli process. In each slot, a user either has a packet to transmit or stays silent. For delay tolerant services, both the inter-arrival time between packets and the packet size may follow any general distributions. The only assumption is that the packet size of delay tolerant services is much longer than that of URLLC services. In the rest of the paper, the tasks of URLLC services and delay tolerant services are referred to as short and long packets, respectively.

\subsection{Achievable Data Rate over Wireless Links}
The users can offload tasks to one of the MEC servers via wireless links. Let $\boldsymbol{\beta}$ be the user association vector with entry $\beta_{m,k}^\xi$ denoting whether the $k$th user is associated with the $m$th AP. If the $k$th user is associated with the $m$th AP, then $\beta_{m,k}^\xi = 1$. Otherwise, $\beta_{m,k}^\xi = 0$. We assume that each user can only offload packets to one of the APs, i.e., $\sum_{m \in \mathcal{M}} \beta_{m,k}^\xi = 1$, where $\mathcal{M} ={1,...,M}$ is the set of indices of APs.

\subsubsection{Achievable Rate for URLLC}
We consider orthogonal frequency division multiple access (OFDMA) systems. The number of subcarriers allocated to the $k$th user is denoted as $N_{m,k}^\xi$. Since the packet size of URLLC services is small, it is reasonable to assume that the bandwidth of $N_{m,k}^{\rm u}$ subcarriers is smaller than the coherence bandwidth and the transmission time is smaller than channel coherence time as well.
Thus, each packet is transmitted over a flat fading quasi-static channel.
If the $k$th user is accessed to the $m$th AP, the achievable rate of the $k$th URLLC user, $k \in \mathcal{K}^{\rm u}$, can be approximated by \cite{Yury2014Quasi}
\begin{align}
	R_{k}^{\rm u} \approx
	\frac{ N_{m,k}^{\rm u} W}{\ln2}
	\left[
	\ln \left(  1+ \frac{  \alpha_{m,k}^{\rm u}  g_{m,k}^{\rm u} P_k^{\rm t, {\rm u}} }{\Phi   N_{m,k}^{\rm u} W N_0 }  \right)
	- \sqrt{  \frac{V_k^{\rm u}}{  T_{\rm s} N_{m,k}^{\rm u} W}  }
	f_Q^{-1}(\varepsilon_k^{\rm d,{\rm u}})
	\right]\;  (\text{bits/s}),\label{R_k^U}
\end{align}
where $W$ is the bandwidth of each subcarrier, $\alpha_{m,k}^{\rm u}$ is the large-scale channel gain, $g_{m,k}^{\rm u}$ is the small-scale channel gain, $P_k^{\rm t,{\rm u}}$ is the transmit power, $\Phi$ is a SNR loss coefficient, which reflects the gap between the achievable rate of practical channel codes and the approximation, $N_0$ is the single-side noise spectral density, $f_Q^{-1}$ is the inverse of Q-function, $\varepsilon_k^{\rm d,{\rm u}}$ is the decoding error probability, and $V_k^{\rm u} = 1- 1\Big/{\left(  1+ \frac{ \alpha_{m,k}^{\rm u}  g_{m,k}^{\rm u} P_k^{\rm t, {\rm u}} }{\Phi N_{m,k}^{\rm u} W N_0}  \right)^2 }$.

%

\subsubsection{Data Rate for Delay Tolerant Services}
For delay tolerant services, the packet size is long, and Shannon's capacity is a good approximation of the achievable rate. If the $k$th user is accessed to the $m$th AP, the ergodic capacity of the $k$th user, $k \in \mathcal{K}^{\rm b}$, can be expressed as
	\begin{equation}
	\mathbb{E}_{g_{m,k}^{\rm b}} \left( R_k^{\rm b} \right)
	= \mathbb{E}_{g_{m,k}^{\rm b}} \left[
	 N_{m,k}^{\rm b} W \log_2\left( 1+\frac{  \alpha_{m,k}^{\rm b} g_{m,k}^{\rm b} P_k^{\rm t, {\rm b}}}{  N_{m,k}^{\rm b} WN_0}\right)   \right] \; (\text{bits/s}),\label{rb}
	\end{equation}
where $\alpha_{m,k}^{\rm b}$ is the large-scale channel gain, $g_{m,k}^{\rm b}$ is the small-scale channel gain, $P_k^{\rm t, {\rm b}}$ is the transmit power.

\subsection{Offloading Policies}
\subsubsection{Offloading Policy of URLLC Services} Considering that feedback from receivers to transmitters may cause large overhead and extra delay, we assume that only $1$ bit CSI is available at each transmitter, which indicates whether the small-scale channel gain is above a certain threshold, $g_k^{\rm{ th,u}}$. If the small-scale channel gain is above the threshold, then the packets are offloaded to the MEC with probability one. Otherwise, the offloading probability is zero. Thus, the overall offloading probability, $x_k^{{\rm u}}$, equals the probability that $g_{m,k}^{{\rm u}} \geq g_k^{\rm{ th,u}}$, i.e.,
\begin{align}
x_k^{{\rm u}} = \Pr\{g_{m,k}^{{\rm u}} \geq g_k^{\rm{ th,u}}\} = \int_{g_k^{\rm{ th,u}}}^{\infty} e^{-g} dg = e^{ -g_k^{\rm{ th,u}} },\label{tloc}
\end{align}
where Rayleigh fading is considered.


\subsubsection{Offloading Policy of Delay Tolerant Services} For each long packet, the transmission duration may exceed the channel coherence time. We consider an offloading policy that does not depend on the current small-scale channel gain. When the $k$th user, $k\in \mathcal{K}^{\rm b}$, has a packet to process, the packet is offloaded to the MEC server with probability $x_k^{\rm b} \in [0,1]$ and processed on the local server with probability $(1-x_k^{\rm b})$.


\subsection{Queueing Model}

\begin{figure}[htbp]
	\vspace{-1.0cm}
	\centering
	\begin{minipage}[t]{0.7\textwidth}
		\includegraphics[width=1\textwidth]{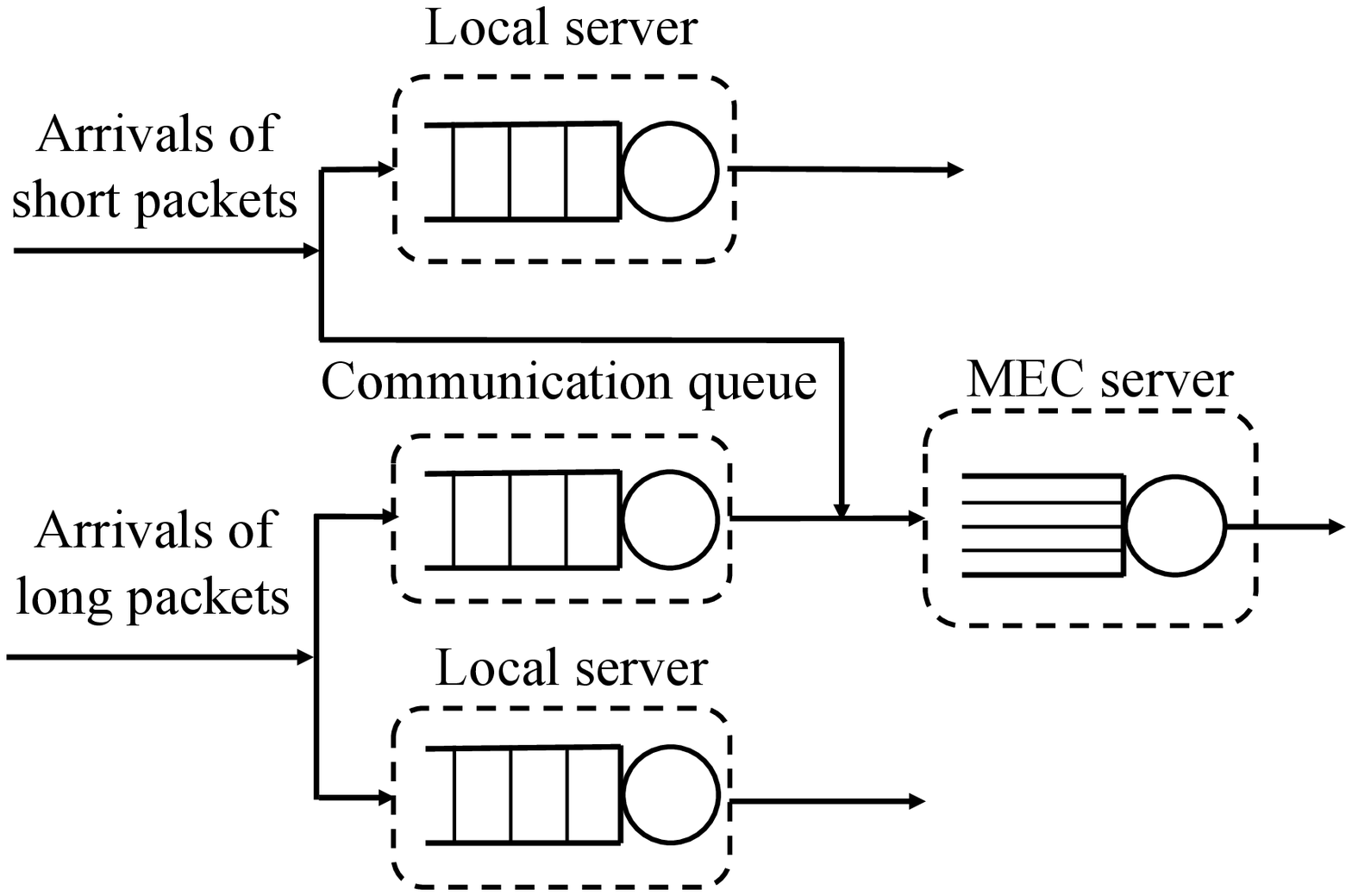}
	\end{minipage}
	\vspace{-1.4cm}
	\caption{Queueing model.}
	\label{mec}
	\vspace{-0.2cm}
\end{figure}

The queueing models of the local servers and the MEC servers are illustrated in Fig. \ref{mec}. In the local servers, packets are served according to the first-come-first-serve (FCFS) order. The difference between URLLC and delay tolerant services lies in the queueing model before uplink transmission. For URLLC services, each packet is transmitted in one slot. Since the packet arrival process follows a Bernoulli process, the peak arrival rate is one packet per slot, which is equal to the transmission rate of the wireless link. As a result, there is no queue before uplink transmission. For delay tolerant services, the peak arrival rate can be higher than the transmission rate, and hence some packets may wait in a communication queue before uplink transmission.

In the MEC servers, there are short and long packets. If the packets are served according to FCFS order, short packets arrive at the MEC servers after a long packet need to wait for the processing of the long packet. To avoid long queueing delay, a processor-sharing (PS) server is adopted at each AP \cite{Mor2013Queue}. On the PS server, the service rate of the server is equally allocated to all the packets in the server. When there are $i$ packets in the $m$th server, the service rate of each packet is $S_m/i$. As shown in \cite{yifan2018delay}, when there are short and long packets, the PS server outperforms the FCFS server.

\subsection{Energy Consumption and Processing Rate at Local Server}
Let $e_k^{\rm loc,\xi}$ be the energy consumption per CPU cycle of the $k$th user. According to the measurements in \cite{Zhang2013Energy,miettinen2010energy}, $e_k^{\rm loc,\xi}=k_0(C_k^\xi)^2$ (J/cycle), where $k_0$ is a coefficient depending on the chip architecture. The typical value of $k_0$ is $10^{-15}$. The energy consumption per packet at the local server is
\begin{equation}
\label{plocc}
E_k^{\rm loc,\xi} = e_k^{\rm loc,\xi} c_k^\xi = k_0(C_k^\xi)^2 c_k^\xi
, \text{(J/packet)},
\end{equation}
which indicates that the energy consumption for processing one packet increases with the processing rate $C_k^\xi$.

\section{Problem Formulation and Deep Reinforcement Learning Framework}
In this section, we first analyze the Quality-of-Service (QoS) constraints of two different services. Then, we formulate an optimization problem to minimize the maximum energy consumption per bit of all the users by optimizing user association, resource allocation, and offloading probabilities subject to the QoS requirements. Finally, we introduce the deep learning framework.

\subsection{QoS Constraints of URLLC Service}
The E2E delay of a packet is defined as the interval between the arrival time of a packet and the time when the processing of the packet is finished. For URLLC service, we denote $D^{\max,\rm u}$ and $\epsilon^{\max,\rm u}$ as the required delay bound and the maximal threshold of the tolerable delay bound violation probability, respectively.


%

\subsubsection{QoS Constraints on Local Servers} If a packet is executed locally, the processing delay is
\begin{equation}
\label{t-loc}
D_k^{\rm{lc,u}}= \frac{ c_k^{\rm u}}{C_k^{\rm u}}\; (\text{slots}).
\end{equation}

When the channel is in deep fading, i.e., $g_{m,k}^{{\rm u}} < g_k^{{\rm th, u}}$, all the packets of a user is served by the local server and the arrival process is a Bernoulli process with average {arrival} rate $\lambda_{k}^{{\rm u}}$. Given a constant service rate, $C_k^{\rm u}$, the queueing model is a Geo/D/1/FCFS model. The complementary cumulative distribution function (CCDF) of queueing delay, $D_k^{{\rm lq, u}}$, in Geo/D/1/FCFS model is given by \cite{Gravey}
\begin{equation}\label{pr}
\begin{split}
\Pr \{D_k^{{\rm lq,u}} > i\} =&
1- \frac{1-(1-x_k^{{\rm u}})\lambda_{k}^{{\rm u}} D_k^{{\rm lc,u}}}{\left( 1-(1-x_k^{{\rm u}})\lambda_{k}^{{\rm u}}\right) ^{i+1}} \sum_{l=0}^{j} \left( (1-x_k^{{\rm u}})\lambda_{k}^{{\rm u}} \left( 1 - (1-x_k^{{\rm u}})\lambda_{k}^{{\rm u}} \right)^{D_k^{{\rm lc,u}}-1}  \right)^l \\
& (-1)^l \binom{i+l-l D_k^{{\rm lc,u}}}{l} ,  \text{if } j D_k^{{\rm lc,u}} \leq i  \leq  (j+1)D_k^{{\rm lc,u}} -1.
\end{split}
\end{equation}

The constraint on E2E delay can be expressed as follows,
\setcounter{equation}{6}
\begin{equation}
\label{dloc}
D_k^{{\rm lc,u}} + D_k^{{\rm lq,u}} \leq D^{\max,{\rm u}}.
\end{equation}
The queueing delay violation probability should satisfy
\begin{equation}
\label{eloc}
\epsilon_k^{{\rm lq,u}} = {\Pr}\{ D_k^{{\rm lq,u}} > (D^{\max,{\rm u}}-D_k^{{\rm lc,u}}) \} \leq \epsilon^{\max,{\rm u}} ,
\end{equation}
which can be computed according to (\ref{pr}).

\subsubsection{QoS Constraints When Offloading to a MEC Server}
When there are long and short packets in a PS server, an accurate approximation of the CCDF of the processing delay of short packets is given by \cite{yifan2018delay},
\begin{equation}
\label{eu}
\epsilon_k^{{\rm mc,u}} =
(\rho_m^{\rm mc}) ^{ \left(  \frac{S_m D_k^{{\rm mc,u}}}{c_k^{\rm u}} -1 \right)  },
\end{equation}
where $\rho_m^{\rm mc}$ is the workload of the $m$th MEC server, defined as follows,
\begin{equation}
\label{rhou}
\rho_m^{\rm mc} =
\frac{  \sum\nolimits_{k \in \mathcal{K}^{{\rm u}} } x_k^{{\rm u}} \lambda_{k}^{{\rm u}} c_k^{{\rm u}}  +   \sum\nolimits_{k \in \mathcal{K}^{\rm b} } x_k^{\rm b}  \lambda_{k}^{\rm b} \bar{c}_k^{\rm b}
}
{  S_m
}  ,
\end{equation}
where $ \bar{c}_k^{\rm b} $ is the average number of CPU cycles required to process a packet of delay tolerant services. The E2E delay of a packet when offloading to the MEC server should satisfy the following constraint,
\begin{align}
\label{t-mec}
1  + D_k^{{\rm mc,u}}  \leq D^{\max,{\rm u}},
\end{align}
where data transmission occupies one slot.


Due to decoding errors and processing delay violation, the overall packet loss probability can be expressed as $\epsilon_k^{\rm u} = 1-(1-\epsilon_k^{{\rm mc,u}}) (1-\epsilon_k^{{\rm d,u}}) \approx \epsilon_k^{{\rm mc,u}} +\epsilon_k^{{\rm d,u}}$, where the approximation is accurate since $\epsilon_k^{{\rm mc,u}}$ and $\epsilon_k^{{\rm d,u}}$ are extremely small.
Then, the constraint on the reliability of the $k$th user can be expressed as, $\epsilon_k^{{\rm mc,u}} +\epsilon_k^{{\rm d,u}}   \leq \epsilon^{\max,{\rm u}}$. We set the upper bound of the decoding error probability and the upper bound of the processing delay violation probability to be equal, i.e.,
\begin{align}
\label{t-mec3}
\epsilon_k^{{\rm mc,u}} \leq \frac{\epsilon^{\max,{\rm u}}}{2} , \epsilon_k^{{\rm d,u}} \leq \frac{\epsilon^{\max,{\rm u}}}{2}.
\end{align}
As shown in \cite{Changyang2018TWC}, setting different packet loss probabilities to be equal leads to minor power loss. By substituting processing delay violation probability in \eqref{eu} into constraint $\epsilon_k^{{\rm mc,u}} \leq \frac{\epsilon^{\max,{\rm u}}}{2}$, we can derive the constraint on the workload as follows,
\begin{align}
 \rho_m^{\rm mc}
\leq
\left(\frac{\epsilon^{\max,{\rm u}}}{2}\right)^
{\left[ \frac{ c_k^{\rm u} }{S_m ( D^{\max,{\rm u}}-1 ) - c_k^{\rm u} }  \right]}
\triangleq \rho_{\rm th} .
\label{rhoth}
\end{align}

\subsection{Stability of Delay Tolerant Services}
For delay tolerant services, we only need to ensure the queueing system is stable, i.e., the average service rate is equal to or higher than the average arrival rate.

\subsubsection{Rate Constraint of Local Servers} To ensure the stability of the queueing system on local servers, we need to guarantee that the processing rate is higher than the average data arrival rate,
\begin{equation}
\label{embblocser}
{C_{k}^{\rm b}} \geq (1-x_k^{\rm b}) \lambda_{k}^{\rm b} \bar{c}_k^{\rm b},  (\text{cycles/slot}).
\end{equation}
Besides, the processing rate should not exceed the maximal computing capacity of the server, $C_k^{\rm b} \leq C_k^{\max,{\rm b}}$.

\subsubsection{Rate Constraint of Wireless Link} To ensure the stability of the communication queue in Fig. \ref{mec}, we need to guarantee that the average transmission rate of the wireless link is equal to or higher than the average data arrival rate, i.e.,
\begin{equation}
\label{embbre}
\mathbb{E}_{g_{m,k}^{\rm b}}\left( R_k^{\rm b} \right)   \geq x_k^{\rm b} \bar{b}_k^{\rm b} \lambda_{k}^{\rm b}  / T_{\rm s},
\end{equation}
where $\bar{b}_k^{\rm b}$ is the average number of bits in a long packet.

\subsubsection{Workload Constraint on the MEC Server}
\label{dtmec}
In the case that only delay tolerant services offload packets to the $m$th MEC server, $x_k^{\rm u} = 0, \forall k \in K^{\rm u}$, the stability of the PS server can be satisfied if the workload meets the following constraint,
\begin{align}
\rho_m^{\rm mc} = \frac{  \sum\nolimits_{k \in \mathcal{K}^{\rm b} } x_k^{\rm b}  \lambda_{k}^{\rm b} \bar{c}_k^{\rm b}
}
{  S_m
} \leq 1.
\label{rho2}
\end{align}
Otherwise, constraint \eqref{rhoth} should be satisfied.

\subsection{Objective Function: Normalized Energy Consumption}
Our goal is to minimize the normalized energy consumption, defined as the energy consumption per bit.

\subsubsection{URLLC Services}
For URLLC services, the circuit power at the local server and the average transmit power for packets offloading are $\lambda_{k}^{\rm u} E_k^{{\rm loc,u}}$ and $\lambda_{k}^{\rm u} P_k^{\rm t,{\rm u}} T_{\rm s}$ (J/slot), respectively. Since the average data arrival rate is $\lambda_{k}^{\rm u} b_k^{\rm u}$ (bits/slot), the {normalized energy consumption} is
\begin{align}
\eta_k^{\rm u} =
\frac{ (1-x_k^{\rm u}) \lambda_{k}^{\rm u} E_k^{{\rm loc,u}} + x_k^{\rm u} \lambda_{k}^{\rm u} P_k^{\rm t,{\rm u}} T_{\rm s} }
{ \lambda_{k}^{\rm u} b_k^{\rm u} }
= \frac{ (1-x_k^{\rm u}) E_k^{{\rm loc,u}} }{  b_k^{\rm u} }+
\frac{ x_k^{\rm u} P_k^{\rm t,{\rm u}} T_{\rm s} }{ b_k^{\rm u} }
\text{ (J/bit)}.\label{ptot}
\end{align}

\subsubsection{Delay Tolerant Services}
If a packet is processed at the local server, the average energy consumption is $E_k^{{\rm loc,b}} = k_0(C_k^{\rm b})^2 \bar{c}_k^{\rm b}$, which is obtained from \eqref{plocc}. Then, the energy consumption per bit is $\eta_k^{{\rm loc,b}} =  E_k^{{\rm loc,b}}/\bar{b}_k^{\rm b}$. If the packet is offloaded to a MEC server, the energy consumption and the average amount of data transmitted in each slot can be expressed as $P_k^{\rm t,{\rm b}} T_{\rm s}$ and $x_k^{\rm b} \lambda_{k}^{\rm b} \bar{b}_k^{\rm b}$, respectively. Then, the energy consumption per bit is $\eta_k^{{\rm mec,b}} = P_k^{\rm t,{\rm b}} T_{\rm s} / x_k^{\rm b} \lambda_{k}^{\rm b} \bar{b}_k^{\rm b}$.
Therefore, the normalized energy consumption of user $k$, $k\in\mathcal{K}^{\rm b}$, can be expressed as follows,
\begin{align}
\eta_k^{\rm b} = (1-x_k^{\rm b})\eta_k^{{\rm loc,b}} + x_k^{\rm b} \eta_k^{{\rm mec,b}} = (1-x_k^{\rm b}) \frac{E_k^{{\rm loc,b}}}{\bar{b}_k^{\rm b}} + x_k^{\rm b} \frac{P_k^{\rm t,{\rm b}} T_{\rm s} }{ x_k^{\rm b} \lambda_{k}^{\rm b} \bar{b}_k^{\rm b}} \; (\text{J/bit}).\label{EB}
\end{align}

\subsection{Optimization Problem}
{To avoid the users with bad channel conditions or high task arrival rates experiencing high energy consumption, we take fairness among all the users into consideration by minimizing the maximal normalized energy consumption of the $K^{\rm u}+K^{\rm b}$ users.} If there is a central control plane that manages user association and resource allocation, the optimization problem can be formulated as follows,
\begin{align}
\mathcal{P}_1:
&\min_{\beta_{m,k}^\xi,P_k^{\rm t, \xi}, N_{m,k}^{\xi}, x_k^{\xi}}
\max_{ k \in \mathcal{K}^{\xi } }
  \eta_k^{\xi}\label{max}\\
\text{s.t. }
& x_k^{\xi} \in [0,1] , \forall k \in \mathcal{K}^{\xi} ,
\label{stx}\tag{\theequation a}
\\
& \sum_{ k\in \mathcal{K}^{\xi} } N_{m,k}^{\xi}   \leq N^{\max}, m = 1,...,M
\label{stN}\tag{\theequation b}
\\
& \sum_{m \in \mathcal{M}} \beta_{m,k}^\xi = 1,
\label{stbeta}\tag{\theequation c}
\\
& {\rho_m^{\rm mc} \leq \begin{cases}
	& 1, \text{if }x_k^{\rm u} = 0, \forall k \in \mathcal{K}^{\rm u}; \\
	& \rho_{\rm th}, \text{otherwise},
	\end{cases}}
\label{strho}\tag{\theequation d}
\\
& C_k^\xi \leq C_k^{\max,\xi}, \forall k \in \mathcal{K}^{\xi},
\label{stC}\tag{\theequation e}
\\
& P_k^{\rm t,\xi} \leq P_k^{\max,\xi} , \forall k \in \mathcal{K}^{\xi},
\label{stP}\tag{\theequation f}
\\
&  \eqref{R_k^U}, \eqref{rb}, \eqref{dloc} , \eqref{eloc}, \eqref{t-mec}, \eqref{t-mec3}, \eqref{embblocser} \text{ and } \eqref{embbre},\nonumber
\end{align}
where $N^{\max}$ is the total number of subcarriers of each AP and $P_k^{\max,\xi}$ is the maximal transmit power of the $k$th user. Constraint \eqref{strho} is obtained from (\ref{rhoth}) and \eqref{rho2}. Since the required transmit power is determined by the bandwidth allocation and the offloading probability, it can be removed from the optimization variables. The relation between the optimal solution and {the inputs, i.e., large-scale channel gains and average task arrival rates, }is denoted as $\pi_1:= \boldsymbol{\alpha}, {\boldsymbol{\lambda}} \rightarrow \boldsymbol{\beta}^*,\boldsymbol{N}^*, \boldsymbol{x}^*$, where $\boldsymbol{\alpha} = ({\boldsymbol{\alpha}}^{\rm u}_1,...,{\boldsymbol{\alpha}}^{\rm u}_{K^{\rm u}},{\boldsymbol{\alpha}}^{\rm b}_1,...,{\boldsymbol{\alpha}}^{\rm b}_{K^{\rm b}})^{\rm T}$,
${\boldsymbol{\alpha}}^{\xi}_k = (\alpha^{\xi}_{1,k},...,\alpha^{\xi}_{M,k})^{\rm T}$,
{$\boldsymbol{\lambda} = ({{\lambda}}^{\rm u}_1,...,{{\lambda}}^{\rm u}_{K^{\rm u}},{{\lambda}}^{\rm b}_1,...,{{\lambda}}^{\rm b}_{K^{\rm b}})^{\rm T}$},
$\boldsymbol{\beta} = ({\boldsymbol{\beta}}^{\rm u}_1,...,{\boldsymbol{\beta}}^{\rm u}_{K^{\rm u}},{\boldsymbol{\beta}}^{\rm b}_1,...,{\boldsymbol{\beta}}^{\rm b}_{K^{\rm b}})^{\rm T}$, ${\boldsymbol{\beta}}^{\xi}_k = (\beta^{\xi}_{1,k},...,\beta^{\xi}_{M,k})^{\rm T}$, $\boldsymbol{N} = ({\boldsymbol{N}}^{\rm u}_1,...,{\boldsymbol{N}}^{\rm u}_{K^{\rm u}},{\boldsymbol{N}}^{\rm b}_1,...,{\boldsymbol{N}}^{\rm b}_{K^{\rm b}})^{\rm T}$, ${\boldsymbol{N}}^{\xi}_k = (N^{\xi}_{1,k},...,N^{\xi}_{M,k})^{\rm T}$, $\boldsymbol{x} = ({{x}}^{\rm u}_1,...,{{x}}^{\rm u}_{K^{\rm u}},{{x}}^{\rm b}_1,...,{{x}}^{\rm b}_{K^{\rm b}})^{\rm T}$, and $(\cdot)^{\rm T}$ denotes the transpose operator.

In practice, a user can subscribe to both kinds of services. If the $k$th user subscribes to both kinds of services, $\lambda_{k}^{\rm u}$ and $\lambda_{k}^{\rm b}$ are referred to as the average task arrival rates of URLLC and delay tolerant services, respectively. The large-scale channel gains of the two kinds of services are the same, i.e., $\alpha_{k}^{\rm u}=\alpha_{k}^{\rm b}$. In the local server of the $k$th user, the packets from different services are waiting in two separated FCFS queues. The energy consumption per packet in (\ref{plocc}) becomes $E_k^{\rm loc} = k_0(C_k)^2 (c_k^{\rm u}+c_k^{\rm b})$. The transmit power constraint of the $k$th user becomes $P_k^{\rm u}+P_k^{\rm d} \leq P_k^{\max}$. The rest of the constraints remains the same.

Note that user association is managed by MME, but resource allocation and offloading probabilities are determined by each AP. The problem $\mathcal{P}_1$ is decomposed into two subproblems that are solved in two timescales at MME and APs, respectively. In the first subproblem, each AP optimizes resource allocation and offloading probabilities with given user association scheme. In the second subproblem, the MME optimizes user association scheme with a DL algorithm, where the behavior of each AP (i.e., the optimal resource allocation and task offloading policy) is taken into account.

\begin{itemize}
	\item Problem $\mathcal{P}_2$: The problem that optimizes subcarrier allocation and offloading probability can be formulated as follows,
	\begin{align}
	\label{p3}
	\mathcal{P}_2: &
	\min_{N_{m,k}^{\xi}, x_k^{\xi}}
	\max_{ k \in \mathcal{K}^{\xi } }
	\eta_k^{\xi},\\
  & \text{s.t.} \;\eqref{stx}, \eqref{stN}, \eqref{strho}, \eqref{stC}, \eqref{stP}, \eqref{R_k^U}, \eqref{rb}, \eqref{dloc} , \eqref{eloc}, \eqref{t-mec}, \eqref{t-mec3}, \eqref{embblocser} \text{ and } \eqref{embbre}. \nonumber
	\end{align}
The relation between the optimal $(\boldsymbol{N}^*,\boldsymbol{x}^*)$ and {$(\boldsymbol{\alpha},\boldsymbol{\lambda},\boldsymbol{\beta})$ is denoted as $\pi_2:= \boldsymbol{\alpha},\boldsymbol{\lambda},\boldsymbol{\beta} \rightarrow \,\boldsymbol{N}^*, \boldsymbol{x}^*$.} The minimal normalized energy consumption achieved with $\pi_2$ is denoted as $Q^*_2(\boldsymbol{\alpha},\boldsymbol{\lambda},\boldsymbol{\beta}|\pi_2)$, which indicates that the normalized energy consumption depends on the user association.
	\item Problem $\mathcal{P}3$: The problem that optimizes user association scheme can be formulated as follows,
	\begin{align}
	\label{p2}
	\mathcal{P}_3: &
	\min_{\beta_{m,k}^\xi}
	Q^*_2(\boldsymbol{\alpha},\boldsymbol{\lambda},\boldsymbol{\beta}|\pi_2), \\
   &\text{s.t. } \eqref{stbeta} \nonumber
	\end{align}
	The relation between the optimal $\boldsymbol{\beta}^*$ and $\boldsymbol{\alpha},\boldsymbol{\lambda}$ is denoted as $\pi_3:= \boldsymbol{\alpha}, \boldsymbol{\lambda} \mathop  \to \boldsymbol{\beta}^*$. The minimal normalized energy consumption achieved with $\pi_3$ is denoted as $Q^*_3(\boldsymbol{\alpha}, \boldsymbol{\lambda}|\pi _2,\pi_3)$, which also depends on $\pi_2$.
\end{itemize}

\subsection{Structure of Deep Learning}\label{DRL}
It is worth noting that both $\mathcal{P}_2$ and $\mathcal{P}_3$ are non-convex. We will propose an optimization algorithm to solve problem $\mathcal{P}_2$ and apply the deep learning algorithm to solve problem $\mathcal{P}_3$.

\vspace{-0.2cm}
\begin{figure*}[ht]
	\centering
	\includegraphics[width=0.7\textwidth]{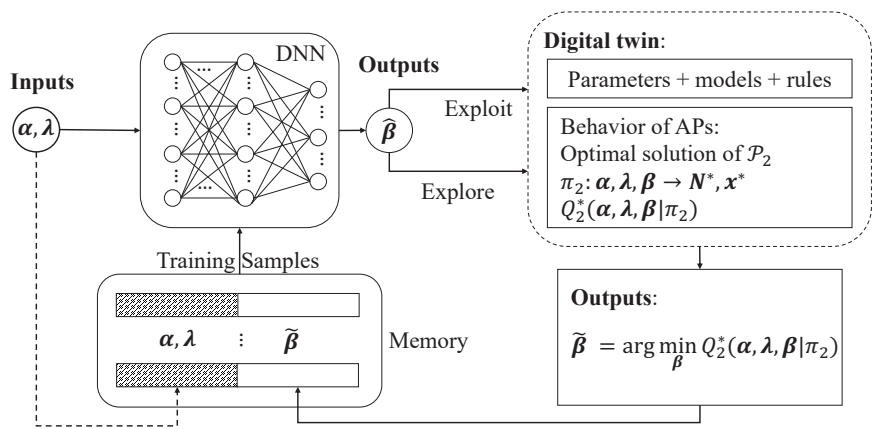}
	\vspace{-0.4cm}
	\caption{{Digital twin enabled DL algorithm.}}
	\label{dnn}
	\vspace{-0.4cm}
\end{figure*}

The {framework of the digital twin enabled DL algorithm} is illustrated in Fig. \ref{dnn}. {The inputs of the DNN are the large-scale channel gains from users to APs and the average task arrival rates of all the users, while the output of the DNN is the user association scheme.} The direct output of the DNN is denoted as $\hat{\boldsymbol{{\beta}}}$, based on which we explore user association schemes. For a user association scheme, we can obtain the related normalized energy consumption from the digital twin. From the feedback of the digital twin, we find the best user association scheme, $\tilde{\boldsymbol{{\beta}}}$, that minimizes the normalized energy consumption among the user association schemes randomly generated according to exploration policies. Finally, the inputs $\boldsymbol{{\alpha}}, \boldsymbol{\lambda}$ and the best output $\tilde{\boldsymbol{{\beta}}}$ are saved in the memory and will be used to train the DNN.


\section{Algorithm to Solve Problem $\mathcal{P}_2$}
In this section, we propose a method to find the optimal solution of problem $\mathcal{P}_2$. Note that when the user association scheme is given, problem $\mathcal{P}_2$ can be decomposed into multiple single-AP problems. In this section, we omit index $m$ for notational simplicity.

\subsection{Outline of the Algorithm}
{From problem $\mathcal{P}_2$ we can see that only constraints \eqref{stN} and \eqref{strho} depend on the optimization variables of all the users, and the other constraints only depend on the resource allocation and offloading probability of a single user. To solve problem $\mathcal{P}_2$, we first remove constraints \eqref{stN} and \eqref{strho} and decompose the problem into multiple single-user problems. After solving these single-user problems, we check whether constraints \eqref{stN} and \eqref{strho} are satisfied or not. The algorithm is summarized in Table \ref{sub}.
}

\begin{table}
	\caption{{Offloading and Subcarrier Allocation Algorithm} }
\vspace{-0.9cm}
	\begin{tabular}{p{16cm}}
		\\\hline
	\end{tabular}
	\label{sub}
	\begin{algorithmic}[1]
		\REQUIRE {Large-scale channel gains, $\boldsymbol{\alpha}$, the user association scheme, $\boldsymbol{\beta}$, and the required searching precisions of normalized energy consumption, number of subcarriers, and offloading probability, $\sigma_{\eta}$, $\sigma_N$, and $\sigma_x$.}
		\STATE Initialize $\eta^{\rm lb} = 0$ and {$\eta^{\rm ub}=\cal E$, where $\cal E$ is the maximal normalized energy consumption when the equalities in constraints \eqref{stC} and \eqref{stP} hold,} $\sigma_{\eta}= 10^{-10}$, $\sigma_N=10^{-3}$.
		\WHILE{ $\eta^{\rm ub} - \eta^{\rm lb} > \sigma_{\eta}$ }		
		\STATE $\eta^{\rm th} = (\eta^{\rm ub} + \eta^{\rm lb})/2$.
		\STATE Initialize $N_k^{\rm  lb, \xi} = 0$, $N_k^{{\rm ub, b}} = N^{\max}$ and $N_k^{{\rm{ub,u}}} = \tilde{N}_{k}^{\rm u}$.
		\FOR {$k \in K^\xi$}
		\WHILE{ $N_k^{\rm b,\xi} - N_k^{\rm lb,\xi} > \sigma_N$ }
		\STATE $N_k^{\rm{th},\xi} = (N_k^{\rm ub, \xi} + N_k^{\rm lb,\xi})/2$.
		\STATE Minimize $\eta_k^\xi$ by optimizing $ \hat{x}_k^{\xi}(N_k^{\rm{th},\xi})$ according to the method in Section \ref{bp}.
		\IF { $\hat{x}_k^{\xi}(N_k^{\rm{th},\xi}) == 0$ }
		\STATE $N_k^{\rm{th}, \xi } = 0$; Break;
		\ELSE
		\IF {$\eta_k^\xi < \eta^{\rm th}$}
		\STATE $N_k^{\rm ub,\xi} = N_k^{\rm{th},\xi}$;
		\ELSE
		\STATE $N_k^{\rm lb,\xi} = N_k^{\rm{th}, \xi}$.
		\ENDIF
		\ENDIF
		\ENDWHILE
		\IF {$ \eta_k > \eta^{\rm th} $}
		\STATE $\eta^{\rm lb} = \eta^{\rm th}$; Break. (problem $\mathcal{P}_2$ is infeasible)
		\ENDIF
		\ENDFOR
		\IF {$\sum_{ k\in \mathcal{K}^{\xi} } N_k^{\rm th, \xi} \leq N^{\max}$ and 	$\rho_m^{\rm mc} \leq  \begin{cases}
				1, & \text{if } \sum_{ k\in \mathcal{K}^{\rm u} } N_k^{\rm u} = 0. \\
				\rho_{\rm th}, & \text{if } \sum_{ k\in \mathcal{K}^{\rm u} } N_k^{\rm u} \neq 0.
				\end{cases}	$}
		\STATE $\eta^{\rm ub} = \eta^{\rm th}$; (problem $\mathcal{P}_2$ is feasible)
		\ELSE
		\STATE $\eta^{\rm lb} = \eta^{\rm th}$. (problem $\mathcal{P}_2$ is infeasible)
		\ENDIF
		\ENDWHILE
		\RETURN {{If $\eta^{\rm th}=\cal E$, the problem is infeasible. Otherwise,} $\eta^*:=\eta^{\rm th}$, $ N_k^{\xi*}:= N_k^{\rm{th},\xi}$ and $x_k^{\xi*}:=\hat{x}_k^\xi(N_k^{\rm{th},\xi})$ }
	\end{algorithmic}
\vspace{-0.5cm}
	\begin{tabular}{p{16cm}}
		\\\hline
	\end{tabular}
\end{table}


{To remove constraint \eqref{stN}, we first find the minimum of the maximal normalized energy consumption via binary search. For a given value of $\eta^{\rm th}$, we minimize the total number of subcarriers that is required to guarantee $\max_{ k \in \mathcal{K}^{\xi } }\eta_k^{\xi} \leq \eta^{\rm th}$, i.e.,}
\begin{align}
\min_{N_{k}^{\xi}, x_k^{\xi}}\;&\sum_{ k\in \mathcal{K}^{\rm u} } N_{k}^{\rm u} + \sum_{ k\in \mathcal{K}^{\rm d} } N_{k}^{\rm d},\label{minN}\\
\text{s.t.} \; &\max_{ k \in \mathcal{K}^{\xi } }\eta_k^{\xi} \leq \eta^{\rm th}, \label{etath}\tag{\theequation a}\\
  &\eqref{stx}, \eqref{strho}, \eqref{stC}, \eqref{stP}, \eqref{R_k^U}, \eqref{rb}, \eqref{dloc} , \eqref{eloc}, \eqref{t-mec}, \eqref{t-mec3}, \eqref{embblocser} \text{ and } \eqref{embbre}. \nonumber
\end{align}
{If the required bandwidth is larger than $N^{\max}$, $\eta^{\rm th}$ cannot be achieved, and the minimum of $\max_{ k \in \mathcal{K}^{\xi } }\eta_k^{\xi}$ is higher than $\eta^{\rm th}$. Otherwise, the minimum of $\max_{ k \in \mathcal{K}^{\xi } }\eta_k^{\xi}$ is lower than $\eta^{\rm th}$. Via binary search, $\eta^{\rm th}$ converges to the minimum of $\max_{ k \in \mathcal{K}^{\xi } }\eta_k^{\xi}$, and the corresponding bandwidth allocation and offloading probabilities are the optimal solution of problem $\mathcal{P}_2$ (See proof in Subsection \ref{ba}.).}


{In the second step, we remove constraint \eqref{strho}, and decompose problem \eqref{minN} into multiple single-user problems. For each single-user problem, we search the minimum number of subcarriers allocated to each user via binary search. For a given value of $N_k^{\rm th,\xi}$, we minimize $\eta^{\xi}_k$ subject to $N_k^{\xi} = N_k^{\rm th, \xi}$, i.e.,}
\begin{align}
\min_{x_k^{\xi}}\; & \eta^{\xi}_k,\label{minetak}\\
\text{s.t.} \; & N_k^{\xi} = N_k^{\rm th, \xi}, \label{Nth}\tag{\theequation a} \\
&\eqref{stx}, \eqref{stC}, \eqref{stP}, \eqref{R_k^U}, \eqref{rb}, \eqref{dloc} , \eqref{eloc}, \eqref{t-mec}, \eqref{t-mec3}, \eqref{embblocser} \text{ and } \eqref{embbre}, \nonumber
\end{align}
{ If $\eta^{\xi}_k \leq \eta^{\rm th}$, then $N_k^{\xi *} \leq N_k^{\rm th, \xi}$. Otherwise, $N_k^{\xi *} \geq N_k^{\rm th,\xi}$ (See proof in Subsection \ref{ba}). Thus, $N_k^{\rm th, \xi}$ either converges the minimum of $N_k^{\xi}$ or $N^{\max}$ (i.e., $\eta^{\xi}_k > \eta^{\rm th}$ even with $N_k^{\xi} = N^{\max}$ ).}

{After obtaining the solutions of the single-user problems, we to check whether constraints \eqref{stN} and \eqref{strho} are satisfied or not in Line 23 of the algorithm in Table \ref{sub}.}

{If constraints \eqref{stN} and \eqref{strho} cannot be satisfied by the end of the binary search, problem $\mathcal{P}_2$ is infeasible and the AP cannot guarantee the QoS requirements of all the users associated with it. In this case, the normalized energy consumption of the user association scheme will be set as infinite in the learning framework in Fig. \ref{dnn}. This user association scheme will not be used to train the DNN since the QoS requirements cannot be satisfied.}

\subsection{Optimal Offloading Probability}\label{bp}
In this subsection, we show how to solve problem \eqref{minetak}. Since the offloading probability depends on bandwidth allocation, we denote the optimal offloading probability as $\hat{x}_k^{\xi }(N_k^{\rm{th,\xi}})$.

\subsubsection{URLLC Services}
\label{urllc}
For URLLC services, the offloading probability is determined by the threshold of small-scale channel gain $g_k^{\rm{th,u}}$. To find the optimal offloading probability, we optimize $g_k^{\rm{th,u}}$ by the following three steps to meet all the constraints in problem \eqref{minetak}.

In the first step, we find the minimal energy consumption per packet at the local server. Since the normalized energy consumption increases with the service rate, we first find the minimal service rate that is required to satisfy the constraints on the E2E delay and the queueing delay violation probability, i.e., \eqref{dloc} and \eqref{eloc}. By substituting $D_k^{\rm lc,{\rm u}} = \frac{c_k^{\rm u}}{ C_k^{\rm u} }$ into \eqref{eloc}, we have $ \epsilon_k^{\rm lq,{\rm u}} = \Pr \{D_k^{\rm lq,{\rm u}} > D^{\max,{\rm u}}- \frac{c_k^{\rm u}}{ C_k^{\rm u} } \} $. From the CCDF of the queueing delay in \eqref{pr}, the minimal service rate can be obtained when $\epsilon_k^{\rm lq,{\rm u}} =  \epsilon^{\max,{\rm u}}$. We denote the minimal service rate that is required to satisfy $D_k^{\rm lq,{\rm u}}$ and $\epsilon_k^{\rm lq,{\rm u}}$ as $C_k^{{\rm u}*}$. According to \eqref{plocc}, the minimal energy consumption per packet at the local servers is $E_k^{\rm loc, \rm{u}*} = k_0(C_k^{\rm{u}*})^2 c_k^{\rm{u}}\;(\text{J/packet})$.

In the second step, we find the minimal value of $g_k^{\rm {th, u}}$ that can satisfy the constraints on decoding error probability and maximal transmit power, i.e., $\epsilon_k^{\rm d,{\rm u}} \leq \frac{\epsilon^{\max,{\rm u}}}{2}$ in \eqref{t-mec3} and \eqref{stP}. The decoding error probability can be obtained from \eqref{R_k^U} by setting $T_{\rm s} R_k^{\rm u}=  b_k^{\rm u}$. Then, the required transmit power that satisfies $\epsilon_k^{\rm d,\rm u} = \frac{1}{2} \epsilon^{\max,\rm u}$ can be expressed as follows,
\begin{equation}
P_k^{\rm t,{\rm u}}
= \frac{1}{g_k^{\rm {th,u}}}\varrho,\label{eq:ptmec}
\end{equation}
where
\begin{equation}
\label{A}
\varrho = \frac{N_k^{\rm {th,u}} W N_0 }{\alpha_{k}^{\rm u} }\times
\left[  \exp \left(
\sqrt{  \frac{1} {T_{\rm s} N_k^{\rm {th,u}} W} } f_Q^{-1}(\frac{\epsilon^{\max,{\rm u}}}{2} )
+\frac{ b_k^{\rm u} \ln 2}{T_{\rm s}  N_k^{\rm {th,u}} W}
\right) - 1\right],
\end{equation}
and the approximation $V_k^{\rm u} \approx 1$ is applied, which is accurate when the receive SNR is higher than $5$~dB \cite{Gross2015Delay,sun2018optimizing}. To satisfy the maximal transmit power constraint, we can obtain the minimal $g_k^{{\rm th,u}}$ by substituting \eqref{eq:ptmec} into $P_k^{\rm t,{\rm u}} = P^{\max, {\rm u}}$, i.e.,
\begin{equation}
{g_k^{\min,{\rm u}}} = \frac{1}{ P^{\max, {\rm u}} }\varrho.\label{eq:mingth}
\end{equation}


In the third step, we derive the closed-form expression of the optimal threshold, $\hat{g}_k^{\rm{th,u}}$, that minimizes the normalized energy consumption. Substituting $x_k^{{\rm u}} = e^{ -g_k^{\rm{ th,u}} }$ and $P_k^{\rm t,{\rm u}}
= \frac{1}{g_k^{\rm {th,u}}}\varrho$ into the expression of $\eta_k^{{\rm u}}$ in \eqref{ptot}, we can derive the derivative of $\eta_k^{\rm u}$ on $x_k^{\rm u}$ as follows,
\begin{equation}
\label{partialpnew}
\eta_k^{{\rm u}'} =
\frac{e^{-g_k^{{\rm th, u}}}}{b_k^{\rm u}} \left( {E_k^{\rm loc,{\rm u}*} } - \frac{\varrho T_{\rm s}}{ g_k^{\rm{ th,u}}} - \frac{\varrho T_{\rm s}}{ \left( g_k^{\rm{ th,u}}\right)^2  }
\right).
\end{equation}
From (\ref{partialpnew}), we can see that the sign of $\eta_k^{{\rm u}'}$ is the same as $f(g_k^{\rm{ th,u}}) \triangleq {E_k^{\rm loc,{\rm u}*} } - \frac{\varrho T_s }{ g_k^{\rm{ th,u}}} - \frac{\varrho T_s }{ \left( g_k^{\rm{ th,u}}\right)^2  }$. When $g_k^{\rm th, {\rm u}} \to 0$, $f(g_k^{\rm{ th,u}}) < 0$. When $g_k^{\rm{ th,u}} \to \infty$, $f(g_k^{\rm th, {\rm u}}) > 0$. Moreover, $f(g_k^{\rm{ th,u}})$ strictly increases with $g_k^{\rm{ th,u}}$. Therefore, $\eta_k^{\rm u}$ first strictly decreases and then strictly increases with $g_k^{\rm{ th,u}}$, and there is a unique solution of $g_k^{\rm{ th,u}}$ that minimizes $\eta_k^{{\rm u}}$ (i.e., $f(g_k^{\rm{ th,u}}) = 0$). The solution of $f(g_k^{\rm{ th,u}}) = 0$ can be derived as follows,
\begin{equation}
\label{gth}
\tilde{g}_k^{\rm{ th,u}} = \frac{1}{2} \left(  \frac{\varrho T_{\rm s}}{E_k^{\rm loc,{\rm u}*} } +
\sqrt{ \left(\frac{\varrho T_{\rm s}}{E_k^{\rm loc,{\rm u}*} } \right)^2 + 4\frac{\varrho T_{\rm s}}{E_k^{\rm loc,{\rm u}*} }  } \right).
\end{equation}
If ${g_k^{\min,{\rm u}}} \leq \tilde{g}_k^{\rm{ th,u}}$, then $\tilde{g}_k^{\rm{ th,u}}$ is the optimal threshold that minimizes $\eta_k^{\rm u}$ subject to the transmit power constraint. Otherwise, since $\eta_k^{\rm u}$ increases with ${g}_k^{\rm{ th,u}}$ in the region $(\tilde{g}_k^{\rm{ th,u}},\infty)$, ${g_k^{\min,{\rm u}}}$ is the optimal threshold. Thus, we have
\begin{equation}\label{optgth}
\hat{g}_k^{\rm{ th,u}} = \max\left\{  g_k^{\min,{\rm u}},  \tilde{g}_k^{\rm{ th,u}} \right\}.
\end{equation}
By substituting $\hat{g}_k^{\rm{ th,u}}$ into \eqref{tloc}, we can obtain the optimal offloading probability, $\hat{x}_k^{\rm u}(N_k^{\rm {th,u}}) =  e^{ -\hat{g}_k^{\rm{th,u}} }$.

\subsubsection{Delay Tolerant Services}
\label{embb}
We apply the binary search to find the optimal offloading probabilities of delay tolerant services that meet the constraints of problem \eqref{minetak}.
Given $N_k^{\rm {th, b}}$, the upper bound of the offloading probability that satisfies the constraints on average data rate and maximal transmit power in \eqref{embbre} and \eqref{stP} can be obtained by substituting $P_k^{\rm {t, b}} = P_k^{\max,{\rm {b}}}$ and $\mathbb{E}_{g_{k}^{\rm b}} \left( R_k^{\rm b} \right)$ in \eqref{rb} into $\mathbb{E}_{g_{k}^{\rm b}}\left( R_k^{\rm b} \right)   = x_k^{\rm b} \bar{b}_k^{\rm b} \lambda_{k}^{\rm b}  / T_{\rm s}$. The lower bound of the offloading probability that satisfies the service rate constraint at the local server in \eqref{embblocser} can be obtained by substituting $C_k^{\rm b} = C_k^{\max,\rm b}$ into ${C_{k}^{\rm b}} = (1-x_k^{\rm b}) \lambda_{k}^{\rm b} \bar{c}_k^{\rm b}$. Let $x_k^{{\rm ub,b}}$ and $x_k^{{\rm lb,b}} $ be the upper and lower bounds of the offloading probability, respectively. If $x_k^{\rm lb,b} > x_k^{\rm ub,b}$, the problem is infeasible, which may happen when the average packet arrival rate $\lambda_{k}^{\rm b}$ is large. When the problem is feasible, to find the optimal offloading probability, $\hat{x}_k^{\rm b}(N_k^{\rm th,b})$ in $[ x_k^{\rm lb,b} , x_k^{\rm ub,b} ]$, we need the following proposition,
\begin{prop}\label{P:1}
\em{$\eta_k^{\rm b}$ in \eqref{EB} is convex in $x_k^{\rm b}$.}
\begin{proof}
See proof in Appendix \ref{App:Prop1}.
\end{proof}
\end{prop}

Then, the optimal offloading probability, $\hat{x}_k^{\rm b}(N_k^{\rm th,b})$, that minimizes $\eta_k^{\rm b}$ can be obtained via binary search.

\subsection{Convergence of the Algorithm}
\label{ba}
In this subsection, we first prove that for a given threshold of the normalized energy consumption, $\eta^{\rm th}$, the algorithm in Table \ref{sub} can find the minimum bandwidth that is required to achieve the threshold (From Line 4 to Line 22 in Table \ref{sub}). To prove it, we only need to prove that the normalized energy consumption decreases with $N_k^{\rm th, \xi}$.
\begin{pro}\label{pro1}
\emph{The minimum of the objective function \eqref{minetak} decreases with $N_k^{\rm th, \xi}$ in the region $[N_k^{\rm ub,\xi} , N_k^{\rm \rm lb,\xi}]$.}
	\begin{proof}
		See proof in Appendix \ref{App:Prop2}.
	\end{proof}
\end{pro}
The above property indicates that the binary search converges to the minimal $N_k^{\rm th, \xi}$ that can guarantee $\eta_k \leq \eta^{\rm th}$, unless it is infeasible (as shown in Line 20 of Table \ref{sub}).

To find out whether problem $\mathcal{P}_2$ is feasible or not, we minimize the total number of subcarriers and see whether it is less than the total number of subcarriers of an AP. Besides, we also need to minimize the total offloading probability and see whether it satisfies the constraint in \eqref{strho}.\footnote{The rest of the constraints are satisfied with the solution of problem \eqref{minetak}.} {The following property indicates that minimizing the offloading probability of the $k$th user is equivalent to minimizing the number of subcarriers allocated to it.}

\begin{pro}\label{P:3}
	\em{The optimal offloading probability $\hat{x}_k^\xi(N_k^{\rm th,\xi})$ increases with $N_k^{\rm th, \xi}$.}
	\begin{proof}
		See proof in Appendix \ref{App:Prop3}.
	\end{proof}
\end{pro}
\vspace{-0.2cm}
Therefore, {by minimizing the sum of the numbers of subcarriers, we also obtained the minimum of the sum of the offloading probabilities.} In other words, both the sum of the numbers of subcarriers and the workload at the MEC server are minimized with the algorithm from Line 4 to Line 22 in Table \ref{sub}. {As a result, problem $\mathcal{P}_2$ is feasible if and only if constraints \eqref{stN} and \eqref{strho} are satisfied with $N_k^{\rm{th},\xi}$ and $\hat{x}_k^\xi(N_k^{\rm{th},\xi})$, i.e., the condition in Line 23 in Table \ref{sub}.}

If problem $\mathcal{P}_2$ is feasible when the normalized energy consumption equals $\eta^{\rm th}$, then $\eta^{\rm th}$ is achievable and the minimal normalized energy consumption $\eta^* \leq \eta^{\rm th}$. Otherwise, $\eta^* > \eta^{\rm th}$. Therefore, with the binary search (i.e., Lines 2,3 and Lines 23 to 27), $\eta^{\rm th}$ converges to $\eta^* $. The coresponding $N_k^{\rm{th},\xi}$ and $\hat{x}_k^\xi(N_k^{\rm{th},\xi})$ converge to the optimal solution $N_k^{\xi*}$ and $x_k^{\xi*}$.

\subsection{Complexity Analysis}
Given the required searching precision of the normalized energy consumption $\sigma_{\eta}$, it takes $\mathcal{O}\left( \log_2 (\frac{\eta^{\rm ub}}{\sigma_{\eta}} )\right) $ steps to obtain the minimum of the maximal normalized energy consumption of all the users. To achieve a target $\eta^{\rm th}$, $(K^{\rm u} +K^{\rm b})\mathcal{O}\left( \log_2 (\frac{N_k^{{\rm ub},\xi}}{\sigma_{N}} )\right) $ steps are needed to obtain the required numbers of subcarriers of $K^{\rm u} +K^{\rm b}$ users, where $\sigma_{N}$ is the required searching precision of the number of subcarriers. For a given number of subcarriers, it takes $\mathcal{O}\left( \log_2 (\frac{1}{\sigma_{x}} ) \right) $ steps to obtain the optimal offloading probability that minimizes the normalized energy consumption of the delay tolerant user in the region $[0,1]$, where $\sigma_{x}$ is the required searching precision of offloading probability. For URLLC services, the optimal offloading probability can be obtained in the closed-form expression in \eqref{optgth}. Therefore, the complexity of the algorithm can be expressed as $ (K^{\rm u} +K^{\rm b}) \mathcal{O}\left( \log_2 (\frac{\eta^{\rm ub}}{\sigma_{\eta}} ) \log_2 (\frac{N_k^{{\rm ub},\xi}}{\sigma_{N}} ) \log_2 (\frac{1}{\sigma_{x}} ) \right)$, which increases linearly with $(K^{\rm u} +K^{\rm b})$.

\section{Deep Learning for User Association}
In this section, we discuss how to explore user association schemes and how to train the DNN.

The set of all the weights and biases of the DNN is denoted as $\Theta = \{\boldsymbol{W}^{[l]},\boldsymbol{b}^{[l]},l=1,...,L_{\rm dnn}\}$, where $L_{\rm dnn}$ is the number of layers, $\boldsymbol{W}^{[l]}$ and $\boldsymbol{b}^{[l]}$ are the weights and the biases in the $l$th layer, respectively. The relation between the input and output of the $l$th layer can be expressed as
\begin{align}
\boldsymbol{Y}^{[l]} = f_{\delta}\left(\boldsymbol{W}^{[l]}\boldsymbol{X}^{[l]} + \boldsymbol{b}^{[l]}\right), \label{layer}
\end{align}
where $\boldsymbol{X}^{[l]}$ and $\boldsymbol{Y}^{[l]}$ are the input and output of the $l$th layer, and the activation function, $f_{\delta}(x)$, is an element-wise operation of a vector. In this work, we use ReLU function as the activation function, i.e., $f_{\delta}(x) = \max(0,x)$.

In each learning epoch, the large-scale channel gains, $\boldsymbol{\alpha}$, {and the average task arrival rates, $\boldsymbol{\lambda}$}, are estimated by the system, and are used to calculate $\hat{\boldsymbol{\beta}}$ from the DNN with parameters $\Theta$. With the output $\hat{\boldsymbol{\beta}}$, user association schemes are generated according to the exploration policies. Then, we find the best user association scheme that minimizes the normalized energy consumption. The pair of inputs {$\boldsymbol{\alpha},\boldsymbol{\lambda}$} and the best user association scheme, denoted as $\tilde{\boldsymbol{\beta}}$, are saved in the memory, and will be used to train the DNN. By the end of the epoch, $N_{\rm t}$ training samples, {$(\boldsymbol{\alpha},\boldsymbol{\lambda},\tilde{\boldsymbol{\beta}})$}, are randomly selected from the memory to train the DNN. After the training, $\Theta$ is updated for the next epoch.

\subsection{Exploitation and Exploration of the DNN}
\label{map}
With ReLU function, the outputs of the DNN are continuous variables, i.e., $\hat{\boldsymbol{\beta}}^{\xi}_k = (\hat{\beta}^{\xi}_{1,k},...,\hat{\beta}^{\xi}_{M,k})^{\rm T}$. We first discuss how to explore user association schemes based on the outputs, and validate the impacts of exploration policies on the normalized energy consumption with simulation.

\subsubsection{Highest Value (Exploitation)} For the $k$th user, a direct way to map the continuous variables $\hat{\boldsymbol{\beta}}^{\xi}_k$ to discrete a user association scheme is to access to the AP with the highest output. We denote the index of the AP with the highest output as $m_k^* = \arg \max_{m\in\mathcal{M}} \hat{\beta}^{\xi}_{m,k}$. Then, ${\beta}^{\xi}_{m_k^*,k}(0) = 1$ and $\hat{\beta}^{\xi}_{m,k}(0) = 0, \forall m \ne m_k^*$. The user association scheme is denoted as ${\boldsymbol{\beta}}(0)$.

\subsubsection{One Step Exploration}
Based on ${\boldsymbol{\beta}}(0)$, we change the association scheme of one of the $K^{\rm u}+K^{\rm b}$ users, while the association scheme of the other users remains the same as ${\boldsymbol{\beta}}(0)$. Since only one user changes the scheme, this method is referred to as one step exploration. With this exploration policy, each user may access to $M-1$ APs, and hence there are $\mu_{\text{OS}}=(K^{\rm u}+K^{\rm b})(M-1)$ possible user association schemes, which are denoted as ${\boldsymbol{\beta}}(1),...,{\boldsymbol{\beta}}(\mu_{\text{OS}})$. Different from ${\boldsymbol{\beta}}(0)$, the $[(k-1)(M-1)+m_k^*]$th of element of ${\boldsymbol{\beta}}[(k-1)(M-1)+m]$ is zero. Besides, if $m<m_k^*$, ${{\beta}}^{\xi}_{m,k}[(k-1)(M-1)+m] = 1$. If $m>m_k^*$, ${{\beta}}^{\xi}_{m+1,k}[(k-1)(M-1)+m] = 1$.

\subsubsection{Random Exploration}
With the random exploration policy, each user randomly selects one of $M$ APs with probability $1/M$. The user association schemes generated with this method are denoted as ${\boldsymbol{\beta}}(\mu_{\text{OS}} + 1),..., {\boldsymbol{\beta}}(\mu_{\text{OS}}+ \mu_{\text{RE}})$, where $\mu_{\text{RE}}$ is the number of schemes generated with the method.

%

\subsection{The DNN Training}
From the $1+ \mu_{\text{OS}} + \mu_{\text{RE}}$ user association schemes, we choose the one that minimizes the normalized energy consumption,
$\tilde{\boldsymbol{\beta}} = \arg \min_{i=0,1,...,\mu_{\text{OS}} + \mu_{\text{RE}}} Q^*_2(\boldsymbol{\alpha},\boldsymbol{\lambda},\boldsymbol{\beta}(i)|\pi_2)$,
and save it in the memory. The memory is empty at the beginning of the first epoch, and the initial values of parameters in $\Theta$ follow a zero-mean normal distribution. When the memory is full, the newly obtained training set, $({\boldsymbol{\alpha}},\boldsymbol{\lambda},\hat{\boldsymbol{\beta}})$, replaces the oldest one.

We adopt the experience replay technique in \cite{mnih2015human} to train the DNN using $N_{\rm t}$ training samples. The parameters in $\Theta$ are updated by using the Adam algorithm \cite{kingma2014adam} to reduce a training loss function, defined as
$L(\Theta) =
-\frac{1}{N_{\rm t}}{ \sum_{ n_{\rm t} =1 }^{N_{\rm t}}
\left[ (\tilde{\boldsymbol{\beta}}_{n_{\rm t}})^{\rm T} \log(\hat{\boldsymbol{\beta}}_{n_{\rm t}})  +
(1-\tilde{\boldsymbol{\beta}}_{n_{\rm t}})^{\rm T} \log(1-\hat{\boldsymbol{\beta}}_{n_{\rm t}})
\right] }.$
When the value of $L(\Theta)$ is below a required threshold, $\sigma_L$, the training phase is finished. After the training phase, MME can use the DNN to calculate user association scheme for any ${\boldsymbol{\alpha}},\boldsymbol{\lambda}$.


\vspace{-0.2cm}	
\section{Simulation Results}

\subsection{Simulation Setup}
\begin{figure}[htbp]
	\vspace{-0.2cm}
	\centering
	\begin{minipage}[t]{0.5\textwidth}
		\includegraphics[width=1\textwidth]{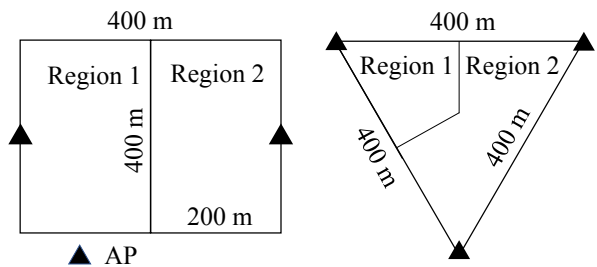}
	\end{minipage}
	\vspace{-0.2cm}
	\caption{Network topologies in our simulation.}
	\label{UD}
	\vspace{-0.2cm}
\end{figure}

The real network topology that will be used in our simulation is illustrated in Fig. \ref{UD}. We vary the user distribution ratio, defined as the user density  in region 1 to the user density in region 2, to see how the impacts of user distribution on normalized energy consumption. The path loss model is $35.3+37.6 \log_{10}(d) $, where $d$ is the distance (meters) between an AP and a user \cite{3GPP2011}. The shadowing is lognormal distributed with $8$ dB standard deviation. The small-scale channel fading follows Rayleigh fading. The packet arrival rate of delay tolerant users, $\lambda_{k}^{\rm b}$, is uniformly distributed between {$5$ and $10$} packets/s.
The packet arrival rate of URLLC users, $\lambda_{k}^{\rm u}$, {is $500$ packets/s \cite{Hou2018Burstiness}}.
Simulation parameters are summarized in Table \ref{tablesys}, unless mentioned otherwise.

The DNN consists of one input layer, four hidden layers, and one output layer, where each hidden layer has {$100$} neurons.
{To achieve a better performance of the DL algorithm, we do not use $\boldsymbol{\alpha}$ and $\boldsymbol{\lambda}$ as the input of the DNN. Instead, the vector $[10\log(\frac{e^{\lambda_1^{\xi}}-1}{\alpha_{1,1}^{\xi}}+1), ...,  10\log(\frac{e^{\lambda_1^{\xi}}-1}{\alpha_{M,1}^{\xi}}+1), ...,
10\log(\frac{e^{\lambda_K^{\xi}}-1}{\alpha_{1,K}^{\xi}}+1), ...,  10\log(\frac{e^{\lambda_K^{\xi}}-1}{\alpha_{M,K}^{\xi}}+1)]^T$ with the size of $M(K^{\rm u} + K^{\rm b}) \times 1$ is used as the input. The element $10\log(\frac{e^{\lambda_k^{\xi}}-1}{\alpha_{m,k}^{\xi}}+1)$ reflects the impacts of $\boldsymbol{\alpha}$ and $\boldsymbol{\lambda}$ on each user's transmit power (dB), which is dominant in the objective function of the normalized energy consumption. The numbers of neurons in the input and output layers are equal to $M(K^{\rm u} + K^{\rm b}) $ and the dimension of $\boldsymbol{\beta}$, respectively.} We set the learning rates of the DNN as $0.001$. The number of training samples in each epoch is $N_{\rm t} = 128$ and the memory can save up to $1024$ training samples. The DL algorithm is implemented in Python with TensorFlow 1.11.

\begin{table}[b]
	\centering
	\caption{Parameters in Simulation}\label{tablesys}
\vspace{-0.2cm}	
	{
		\begin{tabular}{lll}
			\hline
			\multicolumn{1}{|l|}{Notation} & \multicolumn{1}{l|}{Description} & \multicolumn{1}{l|}{Value} \\ \hline
			\multicolumn{1}{|l|}{$S_m$} & \multicolumn{1}{l|}{Computation capability of the $m$th MEC server} & \multicolumn{1}{l|}{$1.6$ GHz} \\ \hline
			\multicolumn{1}{|l|}{$P_k^{\max,\xi}$} & \multicolumn{1}{l|}{Maximal transmit power of each user} & \multicolumn{1}{l|}{$23$ dBm} \\ \hline
			\multicolumn{1}{|l|}{$T_{\rm s}$} & \multicolumn{1}{l|}{Duration of one time slot} & \multicolumn{1}{l|}{$0.125$ms} \\ \hline
			\multicolumn{1}{|l|}{$W$} & \multicolumn{1}{l|}{Bandwidth of each subcarrier} & \multicolumn{1}{l|}{$120$ KHz} \\ \hline
			\multicolumn{1}{|l|}{$N^{\max}$} & \multicolumn{1}{l|}{Maximal number of subcarriers of each AP} & \multicolumn{1}{l|}{$128$} \\ \hline
			\multicolumn{1}{|l|}{$N_0$} & \multicolumn{1}{l|}{Single-sided noise spectral density} & \multicolumn{1}{l|}{$-174$ dBm/Hz} \\ \hline
			\multicolumn{1}{|l|}{$C_k^{\max,\rm b}$} & \multicolumn{1}{l|}{Computation capability of a local server} & \multicolumn{1}{l|}{$5000$ cycles/slot} \\ \hline
			\multicolumn{1}{|l|}{$b_k^{\rm b}$} & \multicolumn{1}{l|}{Number of bytes in a long packet} & \multicolumn{1}{l|}{$[50,100]$ KB} \\ \hline
			\multicolumn{1}{|l|}{$b_k^{\rm u}$} & \multicolumn{1}{l|}{Number of bytes in a short packet} & \multicolumn{1}{l|}{$32$ bytes} \\ \hline
			\multicolumn{1}{|l|}{$k_1$} & \multicolumn{1}{l|}{Number of CPU cycles required to process one byte of information \cite{miettinen2010energy}} & \multicolumn{1}{l|}{$330$ cycles/byte} \\ \hline
			\multicolumn{1}{|l|}{$D_k^{\max,\rm u}$} & \multicolumn{1}{l|}{Delay requirement of URLLC services} & \multicolumn{1}{l|}{$1$ ms} \\ \hline
			\multicolumn{1}{|l|}{$\epsilon_k^{\max,\rm u}$} & \multicolumn{1}{l|}{Maximal tolerable packet loss probability of URLLC services} & \multicolumn{1}{l|}{$10^{-7}$} \\ \hline
		\end{tabular}
	}
\end{table}
\vspace{-0.2cm}

\subsection{Optimal Bandwidth Allocation and Offloading Probabilities}
\label{singleap}
In this subsection, we show the normalized energy consumption achieved by the optimal bandwidth allocation and offloading probability with given user association scheme. In this case, we only need to consider single-AP scenarios. The users are randomly distributed around the AP. Since there is no existing method that optimizes bandwidth allocation and offloading probability for both URLLC and delay tolerant services, we compare the proposed method (with legend `Proposed') with two baselines. In the first baseline, the bandwidth allocation is the same as the optimal solution, but all the packets are offloaded to the MEC (with legend `MEC'). In the second baseline, all the packets are processed at the local servers (with legend `Local'). The normalized energy consumption depends on the location of users and shadowing. In this subsection, we generate $200$ $\boldsymbol{\alpha}, \boldsymbol{\lambda}$ randomly and calculate the average normalized energy consumption.

\begin{figure}[htbp]
	\vspace{-0.2cm}
	\centering
	\begin{minipage}[t]{0.5\textwidth}
		\includegraphics[width=1\textwidth]{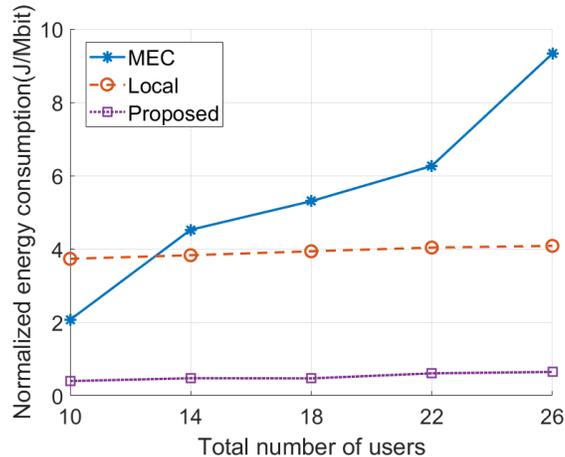}
	\end{minipage}
	\vspace{-0.2cm}
	\caption{Normalized energy consumption v.s. total number of users.}
	\label{maxi-UE}
	\vspace{-0.2cm}
\end{figure}
The normalized energy consumption is shown in Fig. \ref{maxi-UE}, where the number of URLLC users equals the number of delay tolerant users $K^{\rm b} = K^{\rm u}$. The total number of users increases from {$10$ to $26$}. The results show that the normalized energy consumption with the `MEC' scheme increases rapidly as the total number of users increases. The normalized energy consumption of `Local' scheme, however, remains the same as expected. Our proposed scheme can save around $89\%$ of normalized energy consumption compared with the `MEC' scheme and $87\%$ of normalized energy consumption compared with the `Local' scheme.

\subsection{DL Algorithm for User Association}
In this subsection, we show the normalized energy consumptions of different user association schemes, where $N^{\max} = 48$, $S=0.4$~GHz, and $K^{\rm b} = K^{\rm u} = 5$. We compared our DL algorithm (with legend `DL') with the optimal user association scheme (with legend `Optimal') that is obtained by exhaustively searching for all possible user association schemes. {To show the impacts of non-stationary environment on the performance of the DL algorithm, we also provide the performance of a well-trained DNN that will not be updated when the user density varies (with legend `DL Fixed DNN'). With this scheme, there is no exploration and the output of the DNN will be used as the user association scheme.}
Some similar studies focused on offloading and resource allocation with a single AP \cite{you2018asynchronous,Angela2018deep}. The implicit assumption on the user association is that the users are served by the nearest AP or the AP with the highest large-scale channel gain. {In addition, a game theory approach was proposed to optimize resource management and user association in \cite{Jizhe2019Joint}.} Thus, we compared the proposed method with three baselines: With the first baseline method, users are served by the nearest AP (with legend `Nearest AP'). With the second baseline method, users are connected to the AP with the highest large-scale channel gain (with legend `Highest $\alpha$'). {With the third baseline, the game theory approach based on a coalition game in \cite{Jizhe2019Joint} is used to iteratively optimize user association (with legend `Game').
Following \cite{Jizhe2019Joint}, we set the number of the coalitions as $M$ (the number of APs) and the users are randomly chosen to perform Merge, Split and Exchange operations by preferring a smaller objective function as in \eqref{p2}.}

We provide the simulation results in scenarios with different numbers of APs: $M=2$ and $M=3$. When $M=2$, the optimal scheme can be obtained with the exhaustive searching method. However, when $M=3$, the complexity of the exhaustive searching method is too high, and we cannot obtain the optimal scheme.


\vspace{-0.2cm}
\begin{figure*}[htbp]
	\vspace{-0.2cm}
	\centering
	\includegraphics[width=1\textwidth]{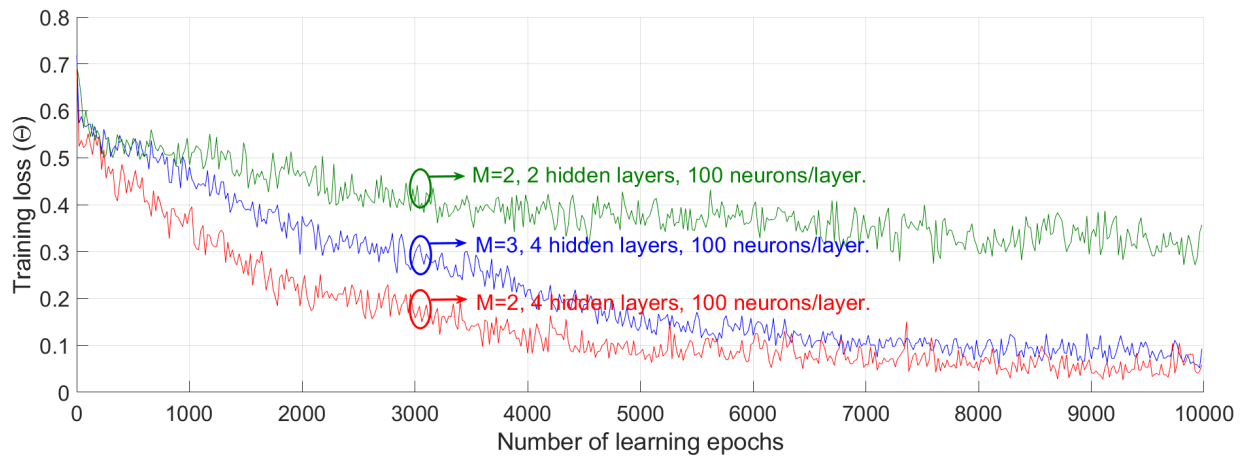}
	\vspace{-1.3cm}
	\caption{{Training loss function v.s. number of learning epoch.}}
	\label{loss23ap}
	\vspace{-0.2cm}
\end{figure*}
To show the convergence of the deep learning algorithm, we provide the values of the training loss function, $L(\Theta)$, as the number of learning epochs increases in Fig. \ref{loss23ap}, where the user distribution ratio is set to be $6:4$. The results with two hidden layers indicate that the DNN does not converge if the number of layers is too small. To find the proper structure of the DNN, we start from the case with one hidden layer and increase the number of layers until the DNN can converge, i.e., four hidden layers. When $M=2$, $L(\Theta)$ is around {$0.1$} after $4000$ epochs. When $M=3$, with $L(\Theta)$ decreases slower than the scenario $M=2$ since the algorithm needs to explore a larger feasible region when $M=3$. For both scenarios, $L(\Theta)$ decreases gradually and approaches to zero.

\begin{figure}[htbp]
	\vspace{-0.2cm}
	\centering
	\begin{minipage}[t]{0.85\textwidth}
		\includegraphics[width=1\textwidth]{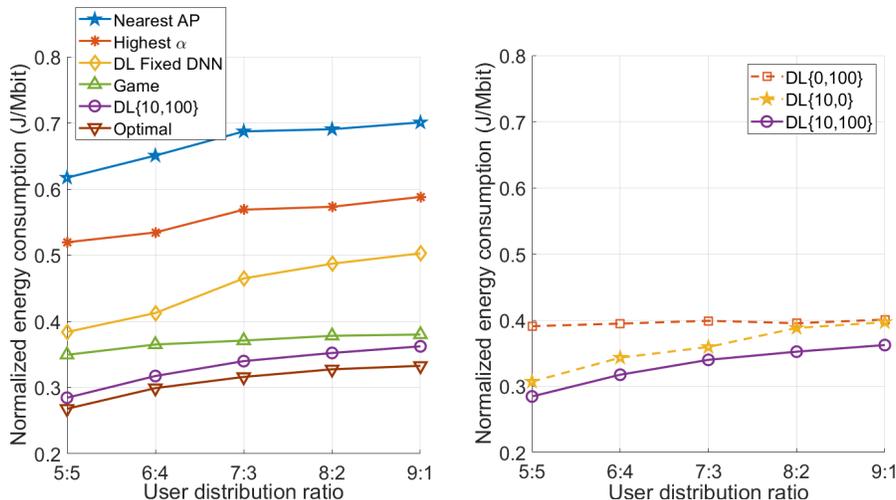}
	\end{minipage}
	\vspace{-0.2cm}
	\caption{{Normalized energy consumption v.s. user distribution ratio, where $M=2$ and $K^{\rm b} = K^{\rm u} = 5$.}}
	\label{distcir}
	\vspace{-0.2cm}
\end{figure}

The averages of the normalized energy consumption in the last $1000$ epochs are shown in Fig. \ref{distcir}, where the numbers of user association schemes generated by the two exploration policies are shown in the legends, e.g., `DL$\{10,100\}$' means $\mu_{\rm OS}=10,\mu_{\rm RE}=100$. The results in Fig. \ref{distcir} show that our proposed method can achieve much smaller normalized energy consumption than the three baseline methods, and perform close to the optimal scheme. {For the `Game' scheme, it converges after $100$ iterations in average. As indicated in \cite{Jizhe2019Joint}, this scheme needs to evaluate the objective function twice in each iteration, which means it needs to evaluate the objective function around $200$ times. However, our proposed algorithm `DL$\{10, 100\}$' only needs to explore ($10 + 100$) user association schemes, i.e., evaluating the objective function $110$ times, which is less than the `Game' scheme. Therefore, our proposed algorithm can achieve a lower normalized energy consumption with less computation complexity. We can also observe that directly exploit the output of DNN without any exploration can save around $30$~\% normalized energy consumption compared with `Highest $\alpha$'. With a few more explorations, the performance can be further improved as shown by `DL$\{10, 100\}$'. Moreover, the one step exploration policy `DL$\{10, 0\}$' can achieve lower normalized energy consumption with less explorations compared with the random exploration policy `DL$\{0, 100\}$'. These results indicate that the output of the DNN can help improving the efficiency of the exploration policy.}

%

\begin{figure}[htbp]
	\vspace{-0.4cm}
	\centering
	\begin{minipage}[t]{1\textwidth}
		\includegraphics[width=1\textwidth]{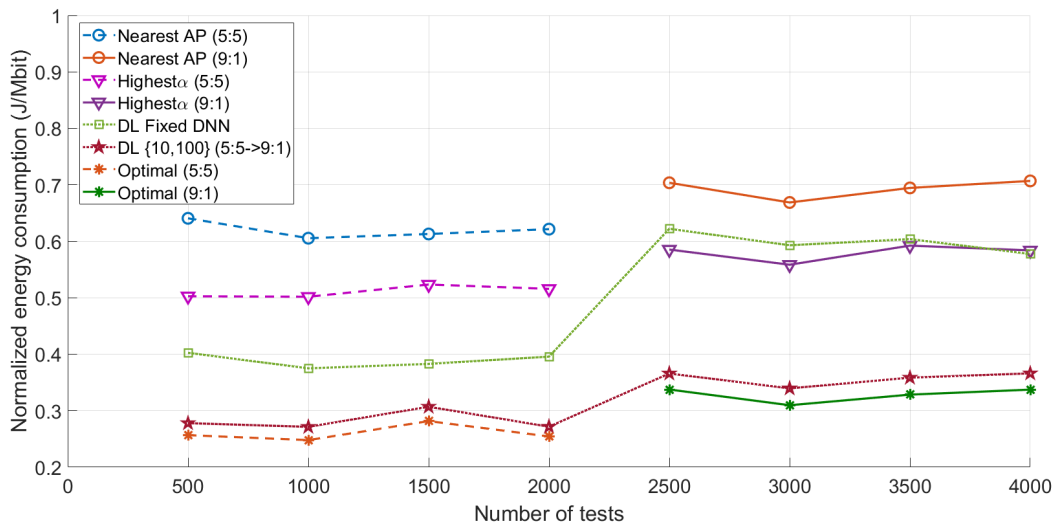}
	\end{minipage}
	\vspace{-0.9cm}
	\caption{{Normalized energy consumption with uncertain user distribution ratios, where $M=2$ and $K^{\rm b} = K^{\rm u} = 5$.}}
	\label{55-91}
	\vspace{-0.2cm}
\end{figure}

In Fig. \ref{55-91}, we study the impacts of {the variation of the user density} on the proposed DL algorithm. In the digital twin, the user distribution ratio is set to be $5:5$. After $1000$ tests with $5:5$ user distribution, the user distribution ratio in the real network becomes different, i.e., $9:1$. The MME needs to update DNN according to the variation of the user distribution. The legends of DL algorithms with the user distribution variation are followed by $(5:5\rightarrow9:1)$. For the other curves, the user distribution ratio is constant. The normalized energy consumptions are the average normalized energy consumptions over $500$ tests. In each test, the large-scale channel gains and the average task arrival rates of the users are generated randomly. {From the results in Fig. \ref{55-91}, we can observe that our proposed DL algorithm `DL$\{10,100\}(5:5\rightarrow9:1)$' can adjust the DNN once the network user distribution ratio changes (after $2000$ tests) and obtain a satisfactory performance compared to the `Optimal$(5:5)$' and `Optimal$(9:1)$'. To further evaluate the importance of the digital twin, we include the legend `DL Fixed DNN' representing that a well-trained DNN is used to make user association decisions without updating its parameters $\Theta$. The results show that when the density of users varies, a fixed DNN can be worse than the baseline method `Highest $\alpha$'. These results indicate that updating DNN according to the non-stationary environment is necessary.}

The normalized energy consumptions achieved with different schemes in the scenario with $3$ APs are provided in Table \ref{perf3ap}. We compare the average of the normalized energy consumptions in the last $1000$ epochs with the `Nearest AP', the `Highest $\alpha$' and the 'DL' schemes. The results in Table. \ref{perf3ap} show that the `DL' scheme can save around $72$~\% and $59$~\% normalized energy consumption compared with the `Nearest AP' and the `Highest $\alpha$' schemes, respectively. This observation indicates that our proposed framework can find an efficient user association scheme when $M=3$.

\vspace{-0.2cm}
\begin{table}[htbp]
	\centering
	\caption{{Performance Comparison When $M=3$ and $K^{\rm b} = K^{\rm u} = 5$}}
	\label{perf3ap}
	\begin{tabular}{|c|c|c|c|}
		\hline
		\text{Schemes}                      & \text{Nearest AP} & \text{Highest $\alpha$} & \text{DL} \\ \hline
		\text{Average normalized energy efficiency (J/Mbit)} & $0.50$              & $0.34$         & $0.14$     \\ \hline
		\text{Normalized energy consumption compared with `Nearest AP'}        & $100\%$              & $68\%$          & $28\%$      \\ \hline
	\end{tabular}
\end{table}
\vspace{-0.4cm}

\vspace{-0.2cm}	
\section{Conclusion}
In this work, we studied how to reduce the normalized energy consumption of users with URLLC and delay tolerant services in a MEC system. We proposed a DL architecture for user association, where a digital twin of network environment was established at the central server for training the algorithm off-line. After the training phase, the DNN was sent to the MME that manages user association. With a given user association scheme, we proposed a low-complexity optimization algorithm that optimized resource allocation and offloading probabilities at each AP. Simulation results indicated that by optimizing resource allocation and offloading probability, our low-complexity algorithm can save more than $87$~\% energy compared with the baselines. Besides, with the DL algorithm, our user association scheme {can achieve lower normalized energy consumption with less computing complexity compared with an existing method and approach to the global optimal solution.}

\appendices
\section{Proof of Proposition \ref{P:1}}
\label{App:Prop1}
\renewcommand{\theequation}{A.\arabic{equation}}
\setcounter{equation}{0}
\begin{proof}
Substituting the equality in (\ref{embblocser}) into (\ref{plocc}), we can obtain that $E_k^{{\rm loc,b}} = k_0 (\lambda_{k}^{\rm b})^2 (\bar{c}_k^{\rm b} ) ^3$. Further substituting $E_k^{{\rm loc,b}}$ into the first term in \eqref{EB}, we can derive the normalized energy consumption at the local server, i.e.,
\begin{equation}\label{eq:A1}
\eta_{k}^{{\rm loc,b}} = (1-x_k^{\rm b}) \frac{E_k^{{\rm loc,b}}}{\bar{b}_k^{\rm b} }  =   \frac{k_0 (\lambda_{k}^{\rm b})^2 (\bar{c}_k^{\rm b} ) ^3}{\bar{b}_k^{\rm b} } (1-x_k^{\rm b})^3  .
\end{equation}
From \eqref{eq:A1}, we can derive that $\frac{\partial^2 \eta_{k}^{{\rm loc,b}}}{\partial (x_k^{\rm b})^2} =
 \frac{6 k_0 (\lambda_{k}^{\rm b})^2 (\bar{c}_k^{\rm b} ) ^3}{\bar{b}_k^{\rm b} }  (1-x_k^{\rm b}) >0$,
and hence the first term in \eqref{EB} is convex in $x^{\rm b}_k$.

To prove the second term in \eqref{EB} is convex in $x_k^{\rm b}$, we only need to prove $P_k^{\rm t,{\rm b}}$ is convex in $x_k^{\rm b}$. From \eqref{embbre}, the required average data rate linearly increases with $x_k^{\rm b}$. Thus, we only need to prove $P_k^{\rm t,{\rm b}}$ is convex in $\mathbb{E}\left( R_k^{\rm b} \right)$. From \eqref{rb} we can see that $R_k^{\rm b}$ increases with $P_k^{\rm t,{\rm b}}$, and it is concave in $P_k^{\rm t,{\rm b}}$. Since the expectation does not change the monotonicity and convexity of the function, $\mathbb{E}\left(R_k^{\rm b} \right)$ is an increasing and concave function with respect to $P_k^{\rm t,{\rm b}}$. According to \cite{Boyd}, if the original function is an increasing and concave function, then the inverse function is a convex function. Therefore, $P_k^{\rm t,{\rm b}}$ is convex in $\mathbb{E}\left(R_k^{\rm b} \right)$.

Since the two terms in \eqref{EB} are convex in $x_k^{\rm b}$, $\eta_k^{\rm b} $ is convex in $x_k^{\rm b}$. The proof follows.
\end{proof}

\section{Proof of Property \ref{pro1}}
\label{App:Prop2}
\renewcommand{\theequation}{B.\arabic{equation}}
\setcounter{equation}{0}
\begin{proof}
For URLLC services, according to \cite{sun2018optimizing}, we know that the required transmit power $P_k^{\rm t,{\rm u}}$ to achieve a certain service rate decreases with $N_k^{\rm th, u}$ when $N_k^{\rm th,u} \leq \tilde{N}_k^{{\rm {u}}}$, where $\tilde{N}_k^{{\rm {th, u}}}$ can be obtained from $\frac{\partial P_k^{\rm t,{\rm u}} }{\partial N_k^{\rm th,u}} = 0$. As a result, $\eta_{k}^{\rm u} = \frac{ x_k^{\rm u}   P_k^{\rm t,{\rm u}} T_{\rm s} }{  b_k^{\rm u} } + \frac{ (1-x_k^{\rm u}) E_k^{\rm loc,{\rm u}} }{  b_k^{\rm u} }$ decreases with $N_k^{\rm th,u}$ when the offloading probability $x_k^{\rm u} \neq 0$ and $N_k^{\rm th,u} \leq \tilde{N}_k^{{\rm {u}}}$.

For delay tolerant services, we need to guarantee the rate constraint of the wireless link in \eqref{embbre}. We substitute $\mathbb{E}_{g_{k}^{\rm b}}\left( R_k^{\rm b} \right) $ in \eqref{rb} into \eqref{embbre}, i.e., $\mathbb{E}_{g_{k}^{\rm b}}\left( R_k^{\rm b} \right)   \geq x_k^{\rm b} \bar{b}_k^{\rm b} \lambda_{k}^{\rm b}  / T_{\rm s}$, and obtain the relationship between $\mathbb{E}_{g_{k}^{\rm b}}\left( R_k^{\rm b} \right) $ and $N_k^{\rm b}$ when the offloading probability $x_k^{\rm b} \neq 0$ as follows
	\begin{equation}
	\mathbb{E}_{g_{k}^{\rm b}} \left[ N_{k}^{\rm b} W \log_2\left( 1+\frac{ \alpha_{k}^{\rm b} g_{k}^{\rm b} P_k^{\rm t, {\rm b}}}{ N_0 N_{k}^{\rm b} W}\right)   \right]
	\frac{T_{\rm s}} {\bar{b}_k^{\rm b}  }
	\geq x_k^{\rm b} \lambda_{k}^{\rm b}.
	\end{equation}
	That is for a given average offloading packet rate $x_k^{\rm b} \lambda_{k}^{\rm b}$, the transmitting power $P_k^{\rm t,{\rm b}}$ decreases with $N_k^{\rm b}$ when the offloading probability $x_k^{\rm b} \neq 0$. As a result, $\eta_{k}^{\rm b} = {P_k^{\rm t,{\rm b}} T_{\rm s} }/{  \lambda_{k}^{\rm b} \bar{b}_k^{\rm b}} + \eta_k^{\rm loc,{\rm b}}$ also decreases with $N_k^{\rm b}$ when the offloading probability $x_k^{\rm b} \neq 0$. This completes the proof.
\end{proof}

\section{Proof of Property \ref{P:3}}
\label{App:Prop3}
\renewcommand{\theequation}{C.\arabic{equation}}
\setcounter{equation}{0}
\begin{proof}
	For the URLLC services, $\varrho$ in \eqref{A} decreases with $N_k^{\rm u}$ in the region $[0,\tilde{N}_k^{\rm u}]$ \cite{sun2018optimizing}. Substituting $\varrho$ in \eqref{A} into the close-form of $\tilde{g}_k^{{\rm th, u}}$ in \eqref{gth}, we can see that $\tilde{g}_k^{{\rm th,u}}$ also decreases with $N_k^{\rm u}$. From $x_k^{{\rm u}} = e^{ -\tilde{g}_k^{\rm{ th,u}} }$ in \eqref{tloc}, we know that $x_k^{\rm u}$ decreases with $\tilde{g}_k^{{\rm th, u}}$. Therefore, $\hat{x}_k^{\rm u}(N_k^{\rm { th, u}}) =  e^{ -\hat{g}_k^{\rm{ th,u}} }$ increases with $N_k^{\rm { th, u}}$, where $\hat{g}_k^{\rm{ th,u}}$ in \eqref{optgth} increases with $\tilde{g}_k^{{\rm th, u}}$.
	
	For the delay tolerant services, to prove the property, we only need to prove that $\eta _k^{\rm b'}$ increases with $\hat{x}_k^{\rm b}$ and decreases with $N_k^{\rm b}$. Hence, when $\eta _k^{\rm b'} = 0$, $\hat{x}_k^{\rm b}$ increases with $N_k^{\rm b}$.

From {Proposition \ref{P:1}}, we know that $\eta_{k}^{\rm b}$ is a convex function in $x_k^{\rm b}$, and hence $\eta_{k}^{{\rm b}'}$ increases with $\hat{x}_k^{\rm b}$.
To prove that $\eta _k^{\rm b'}$ decreases with $N_k^{\rm b}$, we first derive the expression of $\eta_k^{\rm b'}$ as follows,
\begin{align}
\eta_k^{\rm b'} = -\frac{E_k^{\rm loc,{\rm b}}}{\bar{b}_k^{\rm b}} + \frac{ T_{\rm s} }{ \lambda_{k}^{\rm b} \bar{b}_k^{\rm b}} \frac{\partial P_k^{\rm t,{\rm b}}}{ \partial x_k^{\rm b}}, \label{firstorder}
\end{align}
where the expression of $\eta _k^{\rm b}$ in \eqref{EB} is applied. \eqref{firstorder} indicates that to prove $\eta _k^{\rm b'}$ decreases with $N_k^{\rm b}$, we only need to prove that $ \frac{\partial P_k^{\rm t,{\rm b}}}{ \partial x_k^{\rm b}} $ decreases with $N_k^{\rm b}$. By substituting $\mathbb{E}_{g_{k}^{\rm b}}\left( R_k^{\rm b} \right) $ in \eqref{rb} into the rate constraint of the the wireless link in \eqref{embbre}, we can derive that
\begin{equation}
	\label{prop31}
		\frac{\partial x_k^{\rm b}} {\partial P_k^{\rm t,\rm  b}} =
		\frac{ T_s}{ \ln 2 \bar{b}_k^{\rm b} \lambda_{k}^{\rm b} }
		\int_{0}^{\infty} \frac{\alpha_{k}^{\rm b} }{N_0} g_k^{\rm b} e^{-g_k^{\rm b}}  \frac{ 1 }{ 1+
			\frac{ \alpha_{k}^{\rm b} g_k^{\rm b}  P_k^{\rm t,\rm b} } {N_0 W N_k^{\rm b} } } dg.
\end{equation}
Based on \eqref{prop31}, we can see that $\frac{\partial x_k^{\rm b}} {\partial P_k^{\rm t,\rm b}}$ is an increasing function of $N_k^{\rm b}$. According to the characteristic of inverse function (i.e., $ \frac{\partial x_k^{\rm b}} {\partial P_k^{\rm t,\rm b}} \times \frac{\partial P_k^{\rm t,\rm b}}{\partial x_k^{\rm b}} = 1$ at any point $(x_k^{\rm b},P_k^{\rm t,\rm b})$), we can obtain that $\frac{\partial P_k^{\rm t,\rm b}}{\partial x_k^{\rm b}} $ decreases with $N_k^{\rm b}$. As a result, $\eta _k^{\rm b'}$ decreases with $N_k^{\rm b}$.
The proof follows.
\end{proof}


\bibliographystyle{IEEEtran}
\bibliographystyle{unsrt} 
\bibliography{bibfile}

\end{document}